\documentclass[10pt,journal,compsoc]{IEEEtran}

\ifCLASSOPTIONcompsoc
  \usepackage[nocompress]{cite}
\else
  \usepackage{cite}
\fi

\ifCLASSINFOpdf
   \usepackage[pdftex]{graphicx}
\else
   \usepackage[dvips]{graphicx}
\fi

\usepackage{color}
\usepackage[cmex10]{amsmath}
\usepackage{algorithm}
\usepackage{algorithmicx}
\usepackage{algpseudocode}
\usepackage{array}
\usepackage{url}
\usepackage{balance}

\begin{document}

\title{\fontsize{21pt}{\baselineskip}\selectfont Green Internet of Vehicles (IoV) in the 6G Era: Toward Sustainable Vehicular Communications and Networking}



\author{Junhua Wang, Kun Zhu, \emph{Member, IEEE}, and Ekram Hossain, \emph{Fellow, IEEE}  \thanks{J. Wang and K. Zhu are with the College of Computer Science and Technology, Nanjing University of Aeronautics and Astronautics (emails: \{jhua1207, zhukun\}@nuaa.edu.cn). }\thanks{E. Hossain is with the Department of Electrical and Computer Engineering, University of Manitoba (email: Ekram.Hossain@umanitoba.ca).}}

\maketitle

\begin{abstract}
As one of the most promising applications in future Internet of Things, Internet of Vehicles (IoV) has been acknowledged as a fundamental technology for developing the Intelligent Transportation Systems in smart cities. With the emergence of the sixth generation (6G) communications technologies, massive network infrastructures will be densely deployed and the number of network nodes will increase exponentially, leading to extremely high energy consumption. There has been an upsurge of interest to develop the green IoV towards sustainable vehicular communication and networking in the 6G era. However, as a special mobile ad-hoc network, the energy cost in an IoV system involves the communication and computation energy in addition to the fuel consumption and the electricity cost of moving vehicles. Moreover, the energy harvesting technology, which is likely to be adopted widely in 6G systems, will complicate the optimization of energy efficiency in the entire system. Current studies focus only on part of the energy issues in IoV systems without a comprehensive discussion of the state-of-the-art energy-efficient approaches and the  influence of the development of 6G networks on green IoV. In this paper, we present the main considerations for green IoV from five different scenarios, including the communication, computation, traffic, Electric Vehicles (EVs), and energy harvesting management. The literatures relevant to each of the scenarios are compared from the perspective of energy optimization (e.g., with respect to  resource allocation, workload scheduling, routing design, traffic control, charging management, energy harvesting and sharing, etc.) and the related factors affecting energy efficiency  (e.g., resource limitation, channel state, network topology, traffic condition, etc.). In addition, we introduce the potential challenges and the emerging technologies in 6G for developing green IoV systems. Finally, we discuss the research trends in designing energy-efficient IoV systems.
\end{abstract}

\begin{IEEEkeywords}
Green Internet of Vehicles (IoV), 6G, sustainable vehicular communications and networking.
\end{IEEEkeywords}
\IEEEpeerreviewmaketitle

\section{Introduction}
\begin{figure*}[t]
\centering
  \includegraphics[width=0.8\textwidth]{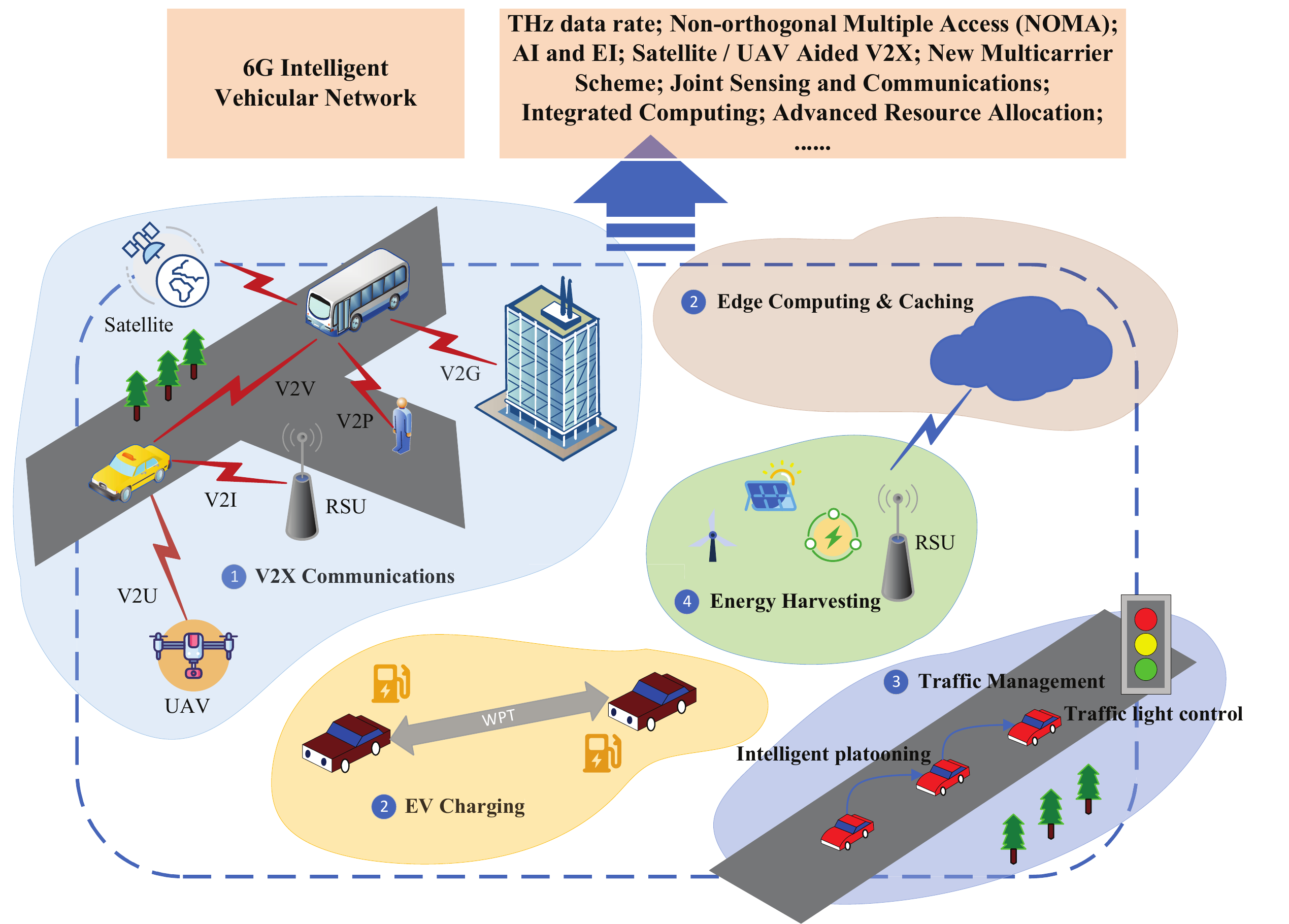}
  \caption{Green IoV scenarios in 6G era}\label{fig:IoV}
\end{figure*}

\IEEEPARstart{A}{s} a fundamental component of the intelligent transportation systems (ITSs), the Internet of Vehicles (IoVs) is developing quickly with the intelligent networked vehicles, the integration of vehicular sensors, and the coordination between vehicles, the road infrastructures and the cloud. As shown in Fig. \ref{fig:IoV}, IoV enables the interaction between vehicles and all the vehicle-related entities via vehicle-to-everything (V2X) communication, which includes the vehicle-to-vehicle (V2V), vehicle-to-infrastructure (V2I), vehicle-to-pedestrian (V2P), vehicle-to-network (V2N), vehicle-to-grid (V2G) and vehicle-to-cloud (V2C) communication, etc. \cite{Ji2020MCOMSTD, Wang2021TVT, Chen2019ACCESS}. Different from a traditional vehicular ad-hoc network (VANET) which mainly focuses on the short-range communication between vehicles and infrastructures, IoV aims to realize wider vehicle information services through different vehicular networking technologies (i.e., onboard information service and heterogenous communication networking) and intelligent technologies (i.e., cognitive computing, big data analysis and artificial intelligence).

IoV communication system can be supported by not only the dedicated-short-range-communication (DSRC) technology, but also the 4G/5G and future 6G technologies with higher data rates and relatively longer transmission distances than that of the DSRC \cite{Zeadally2020MCOMSTD}. From 1G to 6G, the communication rates will improve from kbps to Tera bps, which will promote the development of many new service-oriented applications. In 6G, three types of new services will be provided, including universal mobile ultra-broadband (uMUB), ultrahigh data density (uHDD), and ultrahigh speed low-latency communications (uHSLLC) \cite{Zong2019MVT}. uMUB enables the space-aerial-terrestrial-sea area communication, uHSLLC provides ultrahigh rates and low latency, and uHDD focuses on high-density high-reliability communication. The stringent and diversified requirements in terms of the data rates, data density, latency, reliability and intelligence, will bring more challenges in realizing `green IoV' in the 6G era.

Energy-efficiency will be a critical issue in constructing the 6G-enabled IoV system. First, the increasing number of connected IoV devices and the extensive communication/computation requirements, as well as the growing energy requirement due to the adoption of higher frequency bands in 6G, will lead to surging energy cost in future IoV scenarios. Second, the electricity cost of IoV infrastructures and fuel emissions of vehicles will bring growing energy burdens to IoV system, and make it challenging to develop a sustainable vehicular communication and networking infrastructure. Third, the stringent requirements of quality of service (QoS) and complex intelligent decision algorithms based big data analysis and AI in 6G-enabled IoV applications will cause huge energy consumption and challenge the improvement of energy efficiency.

\setlength{\textfloatsep}{0.25cm}

There has been considerable amount of researches focusing on the realization of energy-efficient vehicular networks. For instance, to improve the energy efficiency in V2X communication scenarios, the joint optimization of channel allocation and communication links selection has been widely studied \cite{Ihsan2021TITS, Saimler2019WCNC, Jameel2021TITS}. To reduce the energy consumption and improve computation utility in vehicular edge computing scenarios, numerous works have been conducted to optimize the offloading decisions, load distribution, and resource allocations, etc. \cite{Dai2020TVT, Huang2021TVT, Guo2020MWC}. In addition, to improve  fuel economy for electric vehicles (EVs), increasing attentions have been paid on choosing the driving routes, development the charging stations, and adjustment of driving behaviors, etc. \cite{Ammous2019TITS, Tang2019JIOT, Ferro2020TVT}. However, these works only focus on the energy-efficiency of a single IoV scenario without a comprehensive discussion of energy consumptions in the entire IoV system. In addition, 6G networks focus on improving the energy-efficiency using low power communication, energy harvesting and energy-efficient computing techniques \cite{Malik2021JIOT}. When considering the realization of `green IoV', it is necessary to take a further step in thinking what challenges the new 6G techniques will bring to the green IoV world. Therefore, in this paper, we consider several critical IoV scenarios, including the V2X communication, vehicular edge computing, traffic management, EV energy management and energy harvesting/sharing, and introduce the energy consumption model under each scenario. We expand the introduction by comparing the related literatures, proposing the unsolved issues and potential researches directions in 6G-enabled IoV systems.

We summarize the main contributions as follows:
\begin{itemize}
  \item We provide a comprehensive review of the energy-efficient IoV techniques from the perspectives of communication networking and computation, traffic management,  energy management of EVs, energy harvesting, etc.
  \item We introduce the most promising 6G-enabled `IoV' scenarios, including the satellite/UAV aided V2X, dynamic energy harvesting, AI-based EV charging decisions, etc.
  \item We discuss the current research challenges for different `green IoV' scenarios, and the emerging issues related to the energy efficiency of IoV in the context of the evolving 6G technologies.
  \item We present the emerging networking technologies that will support the `green IoV' applications, and the potential advantages of a `green IoV' system on the promotion of future network development.
  \item We envision important future research directions for sustainable communication and networking in the 6G era.
\end{itemize}

The rest of the paper is organized as follows. The green IoV scenarios and the green IoV architecture are introduced in Section~\ref{sec:scenarios}. Then we introduce the related researches in detail for different `green IoV' scenarios, including the V2X communication, vehicular edge computing, intelligent traffic management, EVs and charging management, and the energy harvesting management in Section \ref{sec:communication}, \ref{sec:computation}, \ref{sec:traffic}, \ref{sec:evs} and \ref{sec:harvesting}, respectively. We summarize the limitations of existing researches and discuss the future research trend in Section \ref{sec:future}. Section \ref{sec:conclusion} concludes the paper.

\section{Green IoV scenarios and architecture}\label{sec:scenarios}
In the following, we introduce several `green IoV' scenarios including green IoV communication and computing, intelligent traffic management,  energy management of EVs, and energy harvesting and sharing, along with the discussion of current research trends and challenges in each scenario.

\subsection{Green V2X Communications}
The universal connectivity and communication requirements between IoV infrastructures and devices have been drawing significant attentions recently. Increasing communication demands will result in huge energy consumption in deploying IoV infrastructures and transport networks. A cloud-based vehicular radio access network (C-VRAN) architecture can contribute to green vehicular communications by reducing the power consumption of the supporting equipment at base stations, minimizing the interference between remote radio heads (RRHs), shortening the distance between RRHs and users, enabling flexible resource sharing for virtual base stations, and utilizing data depression in the transmission between RRHs and baseband unit (BBU) pool \cite{Su2020MNET}. Besides the architecture design, the optimization of radio resource allocation can also improve the energy efficiency of V2X communication system \cite{Li2019TVT, Guo2019LWC, Zheng2019TVT}. In addition, scheduling optimization can significantly reduce the energy consumption in IoV applications. For example, for battery-enabled road side units (RSUs), the energy consumption can be reduced if the RSU communicates to a vehicle with shorter distance. Another promising method is to enable the RSU or base station to set appropriate service scheduling modes according to the request intensity, so as to reduce the extra energy consumption.

Besides, as a key technology of 6G network, the integrated sensing and communication (ISAC) technology will enhance the communication performance through sensing information \cite{2021Integration, PT2021JCS}. Due to the higher frequency bands, wider bandwidth and denser distribution of antenna arrays used in 6G, the integration of wireless signal sensing and communication will be possible in a single system. In an IoV system, the V2X communication can assist a vehicle in retrieving more traffic information outside of its sensing range, so as to make better driving decisions and save energy consumption with suitable speed adjustment and route planning. For example, in a vehicle platooning system, the multihop V2V communication can enable the last vehicle to sense the driving behaviors of the front vehicle, so as to adjust its driving speed and direction. On the contrary, the sensing information from surrounding environment can facilitate higher-accuracy location, imaging and environment reconstruction, which will improve the V2X communication efficiency with faster interruption recovery and available channel state information (CSI). Therefore, ISAC is a promising technology to enhance the energy-efficiency in future IoV systems.

\subsection{Green Vehicular Edge Computing}
To meet the stringent resource requirements of the computation-extensive tasks in future IoV applications, the edge computing (EC) technology is emerging which can run at the edge of a radio access network and provide computation capacities for IoV devices. The energy utilization of communication and computation can be improved by the edge/cloud computing technologies and the resource scheduling algorithms \cite{Huang2017MCOM, Ning2019TCCN, Ning2019MNET}. However, when vehicles offload computation-intensive tasks to other edge/cloud computing servers, the energy consumptions of transmitting and processing the offloaded tasks increase significantly. In addition, the edge computing servers usually have limited communication, computation and caching capacities, hence, it is challenging to optimize task scheduling and communication/computation resource allocation to reduce the interferences and service delay, and minimize energy consumption in a dynamic vehicular edge computing environment.

\textcolor[rgb]{0.00,0.00,1.00}{}6G networks will use artificial intelligence (AI)-based techniques for resource management, network control and monitoring. The concept of `edge AI' has been proposed recently which pushes the network intelligence at the edge devices, and enables the AI-based learning algorithms to run at distributed edge devices \cite{Saad2020MNET}. The federated learning fits the edge computing architecture well due to the distributed learning model, data privacy protection function and the alleviation of communication overhead \cite{Zhou2021TVT, Huang2021TVT}. However, in a vehicular edge computing network, the mobility of vehicles becomes an issue in keeping continuous information sharing between the edge and the cloud server. Also, the centralized cloud server is vulnerable to security threats, which may lead to a failure of the learning model \cite{Lu2020TII}.

\subsection{Green Intelligent Traffic Management}

Intelligent traffic management involves adaptive control of the traffic lights at intersections, intelligent driving decisions (i.e., speed and distance suggestions), and efficient route planning, etc. A well-designed traffic management solution can facilitate the reduction of energy consumption for both the driving vehicles and roadside infrastructures. For example, a suitable driving speed decision will save the energy consumed in frequent stopping/starting process \cite{USDE2020}.

In 6G-enabled IoVs, the AI-based traffic management strategies are promising to shorten the driven delay and and improve vehicles' energy efficiency \cite{Zhao2016TVT, Shaghaghi2017FITEE, Ge2014TITS}. For driving behavior management, deep learning-based intelligent inter-vehicle distance control algorithm enables a vehicle to determine its speed individually based on the online prediction of communication latency bounds for 6G-enabled cooperative driving with hybrid communication and channel access technologies \cite{Chen2020JIOT}. For intelligent traffic signal control (TSC), the reasonable TSC control method such as deep reinforcement learning (DRL)-based TSC can improve the intersection efficiency. In \cite{Zhong2020MobiWac}, the authors compared four existing methods and the corresponding vehicle speed control strategies, including the Fixed time model (FTM), Single-objective DRL-based TSC method (TSC-SO) which aim at minimizing overall average delay \cite{Genders2016CoRR}, Multiple-objective DRL-based TSC method \cite{Pol2016NIPS}, and Fuel-ECO TSC (FECO-TSC) which is designed to improve vehicle fuel economy under the synthetic scenario and real-world scenario in the City of Toronto \cite{TTS2020}. The authors demonstrate the  performance of the DRL-based TSC algorithm in terms of scheduling efficiency and fuel efficiency. However, to make intelligent management decisions, global traffic information should be collected and updated in real time under the large-scale distributed vehicular network, leading to increasing communication energy consumption and network operation cost. AI-based self-learning and adaptive update algorithm will be attractive in achieving low-complexity intelligent traffic management.

\subsection{Energy Management of Electric Vehicles (EVs)}

The vehicle itself has exhaust emissions, which brings negative effects on human lives and the ecological environment \cite{Wang2020APENERGY, Wu2018JCLEPRO}. There are three main types of vehicles including gas vehicles, electric vehicles (EVs), and hybrid electric vehicles (HEVs). The efficient strategies for saving energy of a gas vehicle mainly consider improving the vehicle's mechanical performance such as its thermal efficiency of the internal combustion engine (ICE) \cite{Zhong2020MobiWac}. The development of EVs or HEVs can reduce Green House Gas (GHG) emission by using clean energy. Based on the data from the US Department of Energy \cite{USDE2020}, the energy loss at the drive system of EVs can reduce to about 20\% when compared with the gasoline vehicles, where more than 60\% of the energy is lost in the form of waste heat. Meanwhile, the design of the kinetic energy recovery system will further improve the energy utilization by recovering electrical energy at the changing of moving states. Different energy consumption models of EVs were presented in \cite{Morlock2020TVT, Qi2018TRD}, and on this basis, various optimization models with respect to the driving paths, charging and discharging behaviors have been designed for improving the EV's energy efficiency \cite{Mahler2014TITS, Zeng2015TCST}.

Another critical issue is to improve the EV charging efficiency. With the coexistence of wired and wireless power transfer technologies, and various charging schemes (i.e., EV-to-grid, EV-to-EV, EV-to-UAV, etc.), the coordinated charging scheduling between multiple charging stations and EVs becomes important. In addition, more attentions should be paid on how to improve the power transmission efficiency, reduce the cost for establishing power transfer system, and reduce the influences of vehicular mobility on the charging efficiency in the wireless power transfer process.

\subsection{Energy Harvesting Management}

Besides vehicles, IoVs also involve the communication entities of road infrastructures and other communication devices, whose energy consumptions can be reduced by optimizing the infrastructure deployment and resource scheduling, and by introducing the renewable energy, etc. For example, as shown in Fig. \ref{fig:IoV}, the wind or solar-powered RSUs can be installed in rural freeway scenarios without connection to the smart grid \cite{Muhtar2013ICC, Atallah2017TVT}. By utilizing the big data based or AI-based power, mobile traffic and vehicle trajectory forecasting techniques, a better matching is possible between renewable energy utilization and workload arrivals over time and space. This can facilitate efficient renewable energy harvesting and delivery between EVs and power edge servers \cite{Zhou2018MCOM}. Meanwhile, since the energy requirements depend on the dynamic traffic demands, collaborative energy management policy for various IoV devices and infrastructures should be considered to achieve higher energy utilization.

In addition, the radio frequency (RF) energy transfer technology can be used to transfer energy between vehicles and RSUs \cite{Atallah2016MWC}. For example, RSUs can sell redundant energy to the EVs passing through RF energy transfer technology when they have sufficient electricity. On the contrary, the RSU can purchase energy from passing EVs when the electricity level is in a lower state. In such a scenario, collaborative energy management between RSUs and EVs can significantly reduce electric power consumption and improve energy utilization \cite{Wang2019MWC}.

\subsection{Green IoV Architecture}

\begin{figure}[t]
\centering
  \includegraphics[width=0.42\textwidth]{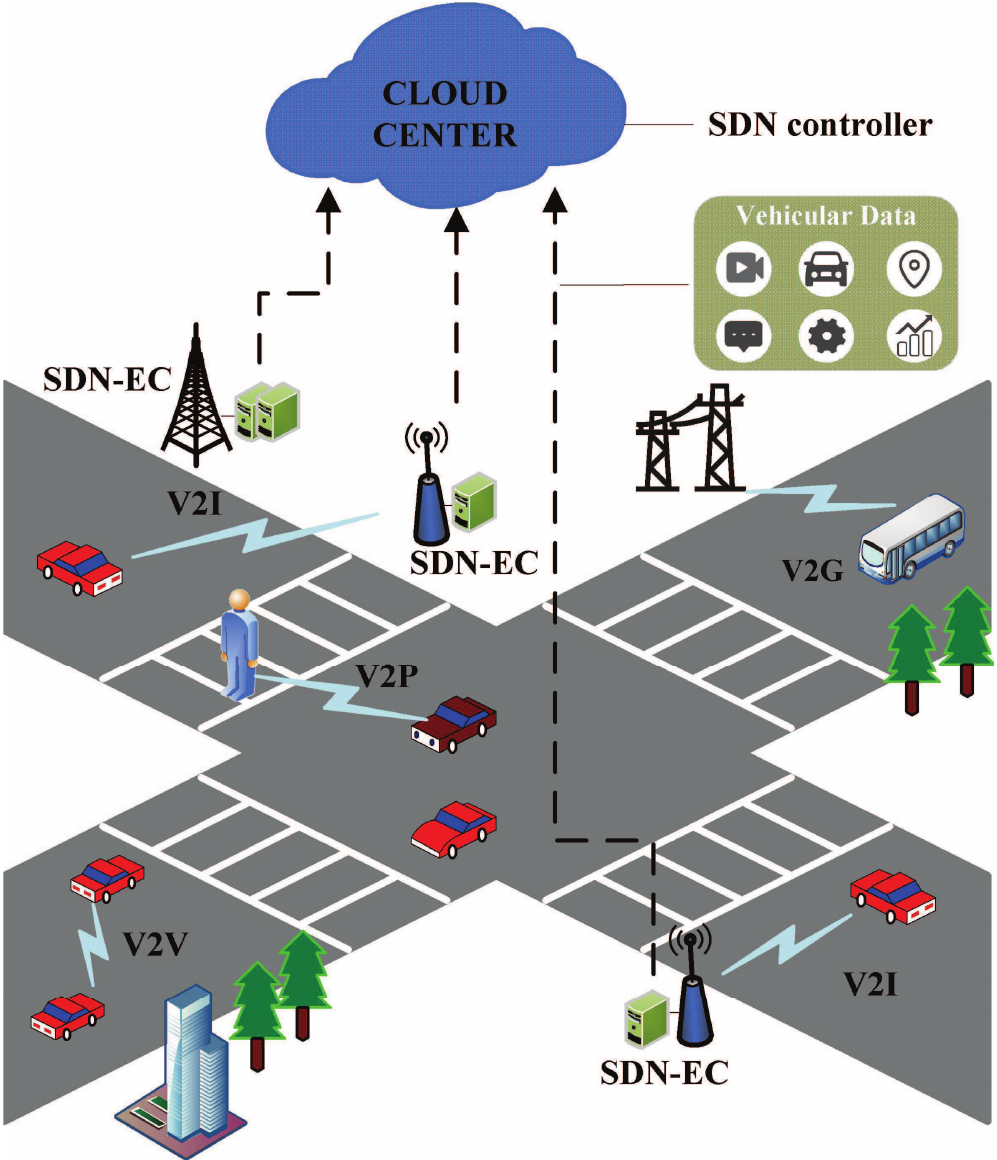}
  \caption{Green SDN-based IoV architecture}\label{fig:architec}
\end{figure}
\setlength{\textfloatsep}{0.25cm}

\begin{figure*}[t]
\centering
  \includegraphics[width=0.8\textwidth]{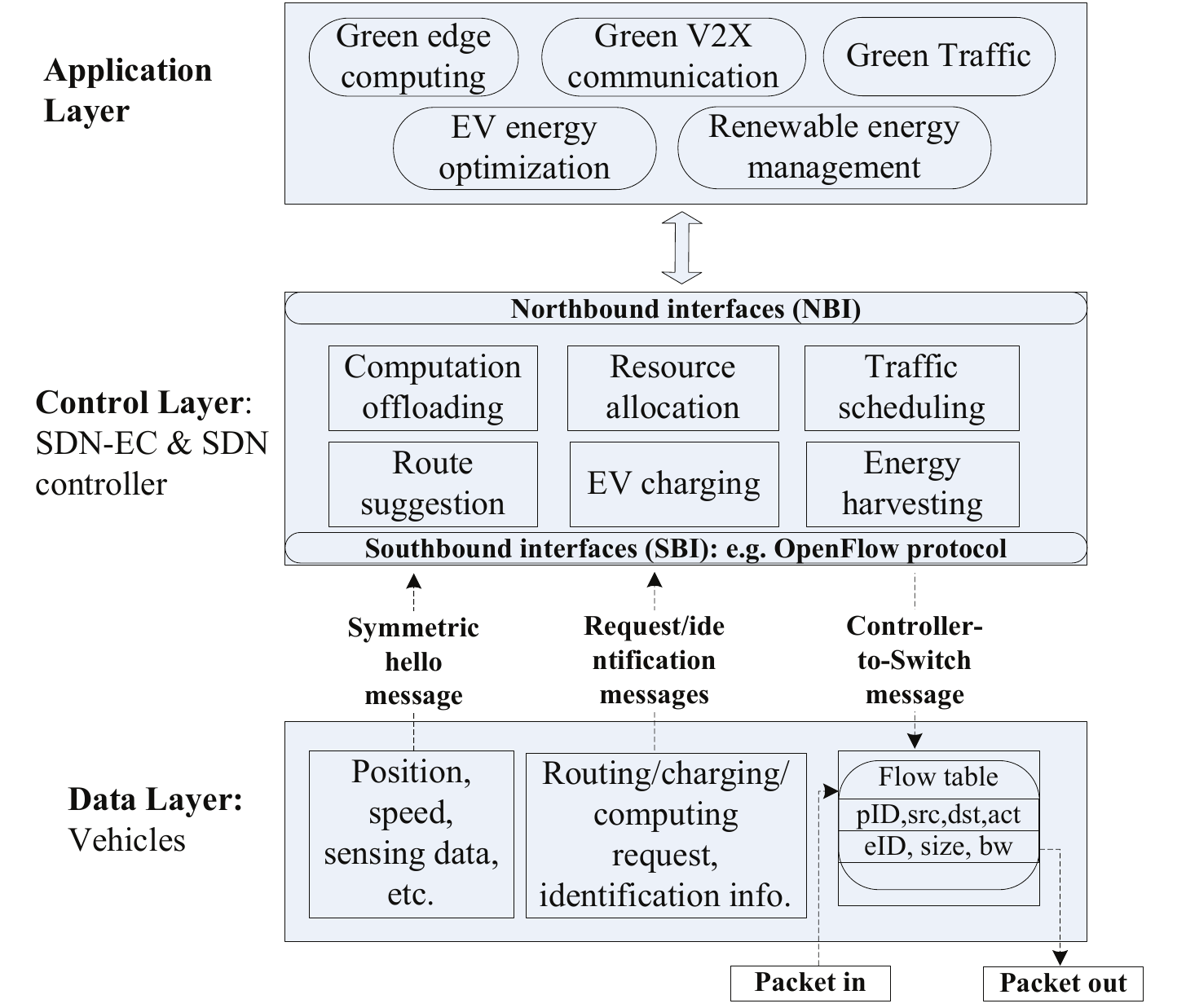}
  \caption{An implementation framework}\label{fig:sdn}
\end{figure*}
\setlength{\textfloatsep}{0.25cm}
Future green IoV systems are expected to be hierarchically integrated with the distributed edge computing components and the remote centralized computing server. As a key enabling technique, the software-defined networking (SDN) philosophy has been widely adopted in designing 5G-enabled vehicular networks. Fig. \ref{fig:architec} shows a system architecture of the green SDN-based IoV. The cloud server acts as the centralized SDN controller. The SDN-enabled edge computing nodes (i.e., SDN-ECs) are distributed to collect vehicular traffic data and request information. If the SDN-ECs can meet the vehicular requests (i.e., computation requests) within expected deadlines, they will complete the requests locally. Otherwise, the SDN-ECs may deliver the requests to other SDN-ECs or the cloud center.

Fig. \ref{fig:sdn} shows an implementation framework of the green SDN-based IoV architecture. It consists of an application layer, a control layer, and a forwarding layer. The application layer provides different green IoV applications, such as the green V2X communication and  green edge computing. The control layer contains the detailed implementation approaches, including the communication/computation resource allocation, intelligent traffic management, etc. The  nodes in the data layer mainly include the vehicles and pedestrians, etc., which may upload traffic data (i.e., position and speed) and request information (i.e., routing and computation requests) to the control layer. After the SDN controller makes the scheduling/routing/computing according to the collected traffic data and requests, the decisions will be delivered to the corresponding network nodes via controller-to-switch messages. The scheduling decisions are stored in the flow table. For example, in the routing scheduling, each routing rule is maintained as a flow entry, which contains the specific request, the source and destination nodes, and the corresponding actions.

\section{Green IoV Communications}\label{sec:communication}

V2X communications play a critical role in enabling the traffic safety, multimedia services, and intelligent driving, etc. In 6G network, Tera bps data rate will be achieved to enable an ultra high-rate wireless bus with ultra reliability for the future intelligent vehicles as required by high-precision positioning and sensing applications, and various real-time video/gaming/3D-interaction services. In addition, the grant free access and Non-orthogonal Multiple Access (NOMA) technology will enable multiple users to utilize non-orthogonal resources concurrently, so as to further support massive connectivity and superior spectrum efficiency in distributed IoV scenarios. However, since the directional beamforming is necessary for THz V2X communications, it is challenging to utilize the advantages of beamforming in highly dynamic traffic environment. In addition, it still lacks well-designed energy-efficient collaboration among multi-radio access technologies (i.e., mmWave and THz) for interference management, channel access and resource allocation, etc. Furthermore, it remains unexplored how to coordinate multiple vehicles for the NOMA transmission and the current orthogonal frequency-division scheme, and allocate resources in distributed V2X networks, so as to achieve efficient energy utilization. Last but not the least, although 6G supports satellite/UAV aided V2X communication with wide coverage and flexible arial base stations, it is still challenging to design reliable and energy-efficient communication schemes between moving UAVs and ground vehicles. Accurate channel modeling and energy-efficient resource allocations methods will be required in space-air-ground integrated network.

Table \ref{table:communication} summarizes the related green IoV communication techniques. In the following, we first introduce the  energy consumption model for V2X communication, then we provide a detailed review of V2X communication mechanisms and present how they achieve high energy-efficiency in different vehicular communication scenarios.

\newcommand{\tabincell}[2]{\begin{tabular}{@{}#1@{}}#2\end{tabular}}

\begin{table*}[t]
  \centering\caption{Survey of green IoV communication techniques}
  \label{table:communication}
\begin{tabular}{|p{3.8cm}|p{6.2cm}|p{2.5cm}<{\centering}|}
\hline
\textbf{Publication} & \textbf{Techniques for green communication} & \textbf{Communication scenario} \\
\hline
\hline
\tabincell{l}{W. Kumar, 2014 \cite{Kumar2014TVT}\\ G. Sun, 2019 \cite{Sun2019TITS}} & Setting sleep modes & \tabincell{l}{V2I \\ V2V} \\
\hline
\tabincell{l}{A. Shrivastava, 2018 \cite{Shrivastava2018AEUE} \\ A. Hammad, 2016 \cite{Hammad2016TVT}} & Optimizing data rates and reducing transmission overhead & V2I \\
\hline
\tabincell{l}{D. Saluja, 2020 \cite{Saluja2020TITS} \\ S. Anbalagan, 2021 \cite{Anbalagan2021JIOT}} & Extending communication coverage & \tabincell{l}{V2I \\ V2X} \\
\hline
\tabincell{l}{M. Azimifar, 2016 \cite{Azimifar2016TVT}\\ S. Vemireddy et al. \cite{Vemireddy2020VC, Vemireddy2021WPC}\\ M. Patra, 2017 \cite{Patra2017TVT}\\ H. Ghazzai, 2019 \cite{Ghazzai2019ICVES}} & Relay-assisted forwarding & V2I \\
\hline
J. Zhao, 2019 \cite{Zhao2019ACCESS} & Optimizing beamforming vectors to reduce interference & V2V \\
\hline
\tabincell{l}{R. Bauza, 2013 \cite{Bauza2013LCOMM}\\ D. Yoo, 2013 \cite{Yoo2013IJDSN}\\ K. Liu, 2016 \cite{Liu2016TITS}} & Reducing retransmissions & V2V \\
\hline
\tabincell{l}{A. Dua, 2015 \cite{Dua2015DSDIS}\\ S. Sattar, 2018 \cite{Sattar2018Comcom}} & Restrictive flooding & V2V \\
\hline
\tabincell{l}{D. Zhang, 2013 \cite{Zhang2013Comjnl}\\ J. Chang, 2014 \cite{Chang2014Computing}\\
K. Satheshkumar, 2021 \cite{Sathesh2021JAIHC}\\ T. Baker, 2018 \cite{Baker2018ICIC}} & Selecting energy-efficient relays & V2V \\
\hline
\tabincell{l}{A. Hammad, 2013 \cite{Hammad2013TVT}\\ Z. Zhou, 2018 \cite{Zhou2018TII}\\ C. Zheng, 2019 \cite{Zheng2019TVT}\\ H. Xiao, 2020 \cite{Xiao2020TITS}\\ Y. Nakayama, 2020 \cite{Nakayama2020GLOBECOM}} & Optimizing resource allocation & \tabincell{l}{V2I\\ V2V\\ V2X} \\
\hline
\tabincell{l}{Z. Ning, 2019 \cite{Ning2019MCOM} \\ Y. Zhang, 2021, \cite{Zhang2021TVT} \\ S. Gu, 2021 \cite{Gu2021JIoT}\\ G. Raja, 2021 \cite{Raja2021TII}\\ T. Limbasiya, 2020 \cite{Limbasiya2020ACM}\\ C. Xu, 2018 \cite{Xu2018TCSVT}\\ H. Wu, 2020 \cite{Wu2020JSAC}\\ H. Wu, 2020 \cite{Wu2020JIOT}\\ G. Qiao, 2020 \cite{Qiao2020JIoT}} & Edge caching strategies & \tabincell{l}{V2I \\ V2X} \\
\hline
\tabincell{l}{P. Dong, 2016 \cite{Dong2016INFCOMW}\\ A. Dua, 2015 \cite{Dua2015DSDIS}\\ N. Kumar, 2020 \cite{Kumar2020TITS}} & Clustering-based routing & V2X \\
\hline
\tabincell{l}{S. Murugan, 2020 \cite{Murugan2020JCR}\\ X. Wang, 2021 \cite{Wang2021TGCN}\\ J. Toutouh, 2013 \cite{Toutouh2013CC}\\ Y. Zeng, 2013 \cite{Zeng2013WN}} & Optimizing routing paths & V2X \\
\hline
P. Sun, 2020 \cite{Sun2020TSUSC} & Optimizing handover operations & V2X \\
\hline
\end{tabular}
\end{table*}
\setlength{\textfloatsep}{0.23cm}

\subsection{Energy Consumption Model}
In V2X communications, the majority of energy consumption comes from the IoV infrastructures and the transmission between IoV devices. As an example, when the BS acts as the IoV infrastructure for providing connectivity to vehicles, we compute the communication energy consumption by focusing on the transmission energy consumptions of vehicles and total energy consumption at the BS. The energy consumption of the BS considers the sleep states and the working state as follows \cite{Mao2021CoRR, Wang2019JASC}:
\begin{equation} \label{EqBS}
{P_{bs}} = {P_{sleep}} + {I_{bs}}\left\{ {{P_{add}} + \eta {P_{trans}}} \right\},
\end{equation}
where $P_{bs}$ and $P_{trans}$ represent the total power consumption and the maximum transmission power consumption of the BS, and $\eta  \in \left[ {0,1} \right]$ represents the usage rate. $P_{sleep}$ is the constant power consumption at the sleep state. $P_{add}$ represents the additional constant power for computation, backhaul communication, and power supply in active mode. $I_{bs}$ is the binary parameter which indicates the sleep or working state of the BS. According to Eq. (\ref{EqBS}), by switching the state of idle BSs to the sleep mode, the energy consumption of the BS can be reduced. In addition, with suitable user association and resource allocation solutions, the transmission energy consumption in active mode can also be reduced.

For uplink communication between a vehicle and a BS, let us denote by $P_v$ and $G_v$ the vehicle transmission power and the channel gain for the link to the BS, respectively. Then the maximum uplink transmission rate can be computed as:
\begin{equation} \label{EqVrate}
{R_v} = B\log \left( {1 + \frac{{{G_v}{P_v}}}{{N + I}}} \right),
\end{equation}
where $B$ is the transmission bandwidth, $N$ and $I$ represent the noise and interference on the channel, respectively. The transmission energy consumptions of the vehicle can be calculated as:
\begin{equation} \label{EqEnergyTrans}
{E_v} = \frac{{{P_v}{D_v}}}{{{R_v}}},
\end{equation}
where $D_v$ is the size of transmission workload.

In the IoV, energy is consumed  to support communications among vehicles, and between vehicles and IoV infrastructures. In addition, energy is consumed to keep the working state of the IoV infrastructures. Therefore, by adjusting the transmission rates, choosing communication time slots, and using edge caching nodes, the  communication efficiency can be improved. For the IoV infrastructure, it is not necessary to always keep the working state and energy can be saved by switching the infrastructure nodes between  sleep and active modes. Furthermore, in multihop vehicular routing, energy efficiency can be improved by controlling the routing hops, selecting suitable relays, etc.

\subsection{V2I Communications}
V2I communications enable vehicles to upload real-time traffic data to RSUs, and also, RSUs can broadcast emergency information to passing vehicles for safety warnings and driving behavior suggestions. However, with the increasing number of users and growing demands for data upload/download, the energy consumptions of both users and RSUs are increasing rapidly. The communication energy efficiency V2I communications can be improved through optimizing the transmission rate, channel and time slot, reducing co-channel interference, setting sleep mode, selecting forwarding relays and utilizing edge caching techniques, etc.

\textbf{\emph{1) Adaptive transmission scheduling:}} When transmitting data to a user with good channel condition, less power is required. Therefore, by estimating the optimum data rate and optimum number of users with good channel conditions and serving the users by multicasting the service at the optimal data rate, the power efficiency and throughput of RSU broadcasting can be improved. The authors in \cite{Shrivastava2018AEUE} consider a scheduling scenario where the RSU is installed to provide infotainment services to the vehicles within the communication range. Each broadcast time slot is equivalent to the coherence time during which the channel state remains unchanged. The optimum SNR value of a channel at which the throughput is maximized or the energy consumption per bit is minimized is calculated. Then the optimum data rate is derived without knowing the channel state information at the RSU. The users with  SNR values higher than the optimum SNR are selected and served with the optimum data rate. Experimental results show that the proposed multicasting algorithm can not only improve energy efficiency but also maximize the throughput. Similarly, the authors in \cite{Hammad2016TVT} propose to use a variable bit rate (VBR) air interface to reduce the downlink RSU energy use, where the transmit power is fixed and changes of channel path loss are compensated for by varying the transmit bit rate. The downlink energy cost is estimated according to the vehicle's communication requirement, location and speed to the RSU. Then, the flow-based modes are introduced to minimize the energy schedule generation, which are solved by the optimal offline and greedy-based online VBR time slot schedules.

Except for the transmission rate adjustment, the adaptive transmission range adjustment and transmission time slot allocation can also reduce RSU energy consumption. For example, in a hybrid vehicular network architecture where both mmWave RSUs (i.e., the RSUs that use mmWave radio access technology) and microwave RSUs are deployed on the roadside to support vehicular communications, the network connectivity, data rate and energy efficiency can be improved with a well-designed association strategy between vehicular nodes and the RSU \cite{Saluja2020TITS}. Similarly, considering that the RSU coverage is affected by the obstacles in city roads such as buildings and trees, the authors in \cite{Anbalagan2021JIOT} propose a Memetic-based RSU (M-RSU) placement strategy to improve the network transmission efficiency and increase the coverage area among IoV devices. The M-RSU placement algorithm uses the received signal strength from a vehicle to determine the coverage area of RSU, thus avoiding signal degradation and improving the overall coverage. In addition, the transmission time slot and packet scheduling is promising to reduce energy consumption of the RSU. The downlink V2I energy communication costs can be reduced through dynamically assigning communication time slot to a vehicle according to the predicted location and velocity \cite{Hammad2013TVT}. The packet-based scheduling and time slot-based scheduling problems are, respectively, formulated as the single-machine job scheduling problem with a tardiness penalty, and the Mixed-Integer Linear Program (MILP) problem. Then, the authors design the Greedy Minimum Cost Flow (GMCF) model to reduce the RSU energy consumption, which obtains the near-optimal scheduling performance.

An attractive energy-saving strategy in cellular networks is to set suitable sleep mechanisms for the BSs or access points (APs) according to the traffic intensity \cite{Peng2014LCOMM, Wu2015COMST, Wu2020TGCN}. In a vehicular network, the energy consumption of RSUs can also be reduced through enabling the sleep modes. W. Kumar et al. \cite{Kumar2014TVT} propose to enable the AP to take queue-length-dependent vacations (switches of sleep mode) to save energy with the sacrifices the quality of service (QoS) in V2I motorway network. They modify the TDMA-based protocol as a slot-based packet reservation multiple-access (M-PRMA) protocol, where the queue of time slots is considered as servers and the outage of a slot represents a server on vacation. Under this assumption, the authors study proactive and reactive random sleep strategies. Experimental results show that the introduction of sleep strategies at an AP can save about 80\% of transmission energy during off-peak hours.

\textbf{\emph{2) Relay-assisted scheduling and edge caching:}} The number of hops in downlink V2I transmission may affect the energy consumption, which however, is not the unique factor. Even though the consumed communication energy  increases with more forwarding operations, the multi-hop short-distance forwarding may save more energy than the single-hop long-distance transmission. In particular, with single-hop V2I communication,  vehicles are served by the RSU or BS, which can cause inefficient resource utilization for the links among vehicles and also heavy communication burden at the RSU or BS. Therefore, by combining the vehicle-to-vehicle (V2V) forwarding in the downlink V2I communication, the RSU energy consumption may be reduced. M. Azimifar et al. \cite{Azimifar2016TVT} propose an energy-efficient V2I traffic scheduling method by enabling the RSU to dynamically forward packets through vehicles in the energy-favourable locations. In the offline scheduling, the sequence of arriving vehicles and their communication requirements are assumed to be known in advance, and a scheduling algorithm is proposed to derive the lower bound on the RSU energy to fulfill the vehicular packet communication requirements. In the online scheduling, the authors propose two scheduling algorithms to make causal downlink transmission decisions. One algorithm utilizes a greedy local optimization to create schedules, and the other one uses a finite-window group optimization (FWGO) by grouping vehicles together for joint traffic scheduling. Simulation results show that the proposed algorithm can improve the downlink energy requirements with V2V-assisted packet forwarding.

In a similar spirit, S. Vemireddy and R. Rout in \cite{Vemireddy2020VC} and \cite{Vemireddy2021WPC} propose  energy-efficient cooperative relay scheduling strategy for transmitting data to the target vehicles which are moving outside the RSU's coverage. Various relay selection algorithms are proposed, including the Minimum Cost Flow (MCF), cluster-based approach, and the greedy algorithm, to select vehicles which are in energy-favorable locations and use less time to reach the target vehicle. In \cite{Vemireddy2020VC}, the adopted V2V communication channel is independent of the downlink V2I transmission channel. In \cite{Vemireddy2021WPC}, they further propose a forward relay scheduler (FRS) based on auction theory to optimally assign the relay vehicles with different time slots for minimizing the RSU energy consumption. Experimental results show that the proposed algorithms can improve the energy-efficiency of the RSU and the data delivery ratio to target vehicles in the uncovered area.

In addition, with the edge caching techniques, vehicles can download contents from the  edge  which facilitates the improvement of end-to-end energy efficiency. The authors in \cite{Ning2019MCOM} focus on the selection of edge devices (EDs) for providing backup-resource. They put forward a Green and Sustainable Virtual Network Embedding (GSVNE) framework to study the available number of backup EDs and embed virtual networks onto the suitable EDs, which can serve as many VNs as possible and provide an efficient sharing of network backup resources. The authors in \cite{Zhang2021TVT} focus on resource allocation optimization for maximizing energy-efficiency in cache-based dense vehicular networks with THz communication links. Multiple cache access points (cAPs) are deployed on the roadside for providing vehicles with high-rate data transmission. The authors formulate a joint subband and power allocation problem with the mean-field approximation (MEA) method. Then each cAP is able to solve the local energy-efficiency optimization problem by Dinkelbach method and Lagrangian dual method. Further, a global mean-field game (MFG)-based algorithm is proposed to associate the local optimization strategy of each cell and maximize the energy-efficiency in the cache-based THz vehicular networks.

\subsection{V2V Communications}
The communication energy-efficiency of one-hop or multi-hop V2V transmission has been paid wide attentions in the past few years. T. Darwish et al. \cite{Darwish2017CEE} give a review about green vehicular geographical routing. Several green vehicular routing methods are introduced with objectives such as preventing packet failure and retransmission, reducing routing overhead due to control packets, and reducing the number of hops of packet transmission, etc. However, this survey only considers V2I/V2V-based geographical routing without considering 5G/6G-enabled green V2X communication, where new technologies such as the heterogenous access network and edge caching bring both challenges and potentials in improving communication energy efficiency. In the following, we review some important works in the area of energy-efficiency optimization in V2V communications.

\textbf{\emph{1) Interference management:}}  To improve energy-efficiency in heterogeneous intelligent connected vehicles (ICV) networks, J. Zhao et al. \cite{Zhao2019ACCESS} study a heterogenous access model in V2V communication, where a direct link and a dual-hop relaying link are supported simultaneously. The dual-hop communication is assisted by a dedicated relaying vehicle with multiple antennas. The interferences between two data streams are tackled by jointly designing the receive beamforming and transmit beamforming vectors at the dedicated relaying vehicle. The authors formulate the joint optimization problem to maximize the system energy efficiency, and decompose it into three subproblems with only one unknown beamforming vector for each subproblem. Finally, an iterative algorithm is used to obtain the solutions of three subproblems. Experimental results show the energy-efficiency performance of the proposed beamforming vector optimization algorithm.

\textbf{\emph{2) Reducing transmission/retransmission overhead:}}  The most trivial solution to disseminate data among highly dynamical vehicles is the blind flooding which wastes large amount of energy and bandwidth  due to unnecessary transmission \cite{Dua2015DSDIS}. In a multi-hop routing process, restrictive flooding is more energy-efficient than plain flooding under the same reliability constraint \cite{Sattar2018Comcom}. By limiting the number of relaying nodes and using better routing directions, the energy consumption in end-to-end routing process can be greatly reduced.

The vehicular routing performance strongly depends on the selection of reliable relaying nodes. With reliable multi-hop forwarders, the transmission energy consumption from the source node to the destination node is reduced due to the decreased retransmission operations. To identify the reliable relaying nodes in VANETs, the authors in \cite{Bauza2013LCOMM} use the periodical beacon reception rates from the neighboring vehicles to estimate the quality of a transmission link, and then select the more reliable transmission links to forward packets, so as to reduce retransmission overhead and improve energy efficiency. Another promising approach to reduce the number of retransmissions is to use the network coding technique, which enables multiple vehicles that have common interests in certain data to collaboratively download data with reduced number of downloading operations \cite{Zhu2015TVT, Liu2016TITS, Liu2016TVT}. For example, K. Liu et al. \cite{Liu2016TITS} adopt the network coding techniques to reduce the number of packet transmissions in V2I/V2V communications for improving the network throughput and system energy efficiency. In addition, once retransmission has occurred, the communication energy consumption can be reduced by selecting the relaying nodes with the shortest forwarding delay to rebroadcast the data. The forwarding delay of a vehicle can be calculated by using the vehicle's velocity, distance and angle from nearby vehicles \cite{Yoo2013IJDSN}.

\textbf{\emph{3) Adaptive transmission scheduling:}} By enabling the adaptive employment scheme or `work' mode on the relaying vehicles, they can provide varying computation resources according to the dynamic communication requirements, which will save the energy consumption of redundant forwarding operations. Due to the natural characteristics of long parking time and wide distribution, the parked vehicles have been employed as road relaying nodes to provide communication connectivity for moving vehicles that are far apart or obscured by obstacles. Considering that energy exhaustion may occur when parked vehicles continuously work with no restrictions, the author in \cite{Sun2019TITS} propose the energy-saving method for the parked vehicles which act as relay nodes for providing services between driving vehicles. They classify moving vehicles into different clusters according to their communication coverage and only select certain parked vehicles to enter into the `working' mode for energy saving. For reducing the energy consumption of parked vehicles, the authors use a Markov model for the energy consumption of a parked vehicle  and propose a dynamic work mode selection algorithm.

The relay selection problem has been paid wide attention in designing energy-efficiency vehicular routing protocols \cite{Zhang2013Comjnl, Sathesh2021JAIHC, Chang2014Computing, Baker2018ICIC}. The factors related to relay selection  \cite{Zhang2013Comjnl} mainly include the vehicular categories information (i.e., public vehicle and private vehicle), the driving state of its neighboring vehicles, the distance between the current section and the next intersection. In \cite{Sathesh2021JAIHC}, the relaying vehicles are selected based on vehicle position, directional flow and message delivery time. The vehicle direction (VD)-based authorization selection model can decrease the unnecessary message broadcasting in vehicular routing protocol, thus reducing communication overhead and improving energy efficiency. Similarly, in \cite{Chang2014Computing}, the authors leverage the information of vehicle driving direction, traffic density and the distance between vehicles to select energy-efficient relaying vehicles. The proposed geographical routing algorithm is shown to have higher effectiveness compared with the ad hoc on-demand distance vector and the dynamic source routing (DSR) algorithms. The authors in \cite{Baker2018ICIC} propose to select the routing paths based on the total power consumption of the relaying vehicles between the source and destination nodes. In the proposed routing protocol (i.e., GreeAODV), the power consumption is calculated when the relaying node receives a packet and transmits an incoming packet.

The joint optimization of relay selection and transmission resource allocation is an attractive approach to reduce energy consumption in vehicular networks.  By selecting the energy-efficient relaying vehicles and allocating spectrum and power as required by the relaying vehicles, both the communication quality and energy-efficiency are improved. Z. Zhou et al. \cite{Zhou2018TII} consider to utilize a cooperative two-hop device-to-device based vehicle-to-vehicle (D2D-V2V) transmission to offload high volumes of vehicular data from cellular infrastructures to vehicular networks, and formulate the joint optimization of relay selection, spectrum allocation and power control problem from the energy efficiency perspective. They propose a two-stage energy-efficient resource allocation algorithm, which include an auction-matching based algorithm to jointly optimize  relay selection, spectrum allocation and power control for maximizing energy-efficiency of two-hop D2D-V2V and cellular links in an iterative process.

\subsection{Hybrid V2X Communications}

\begin{figure}[t]
\centering
  \includegraphics[width=0.49\textwidth]{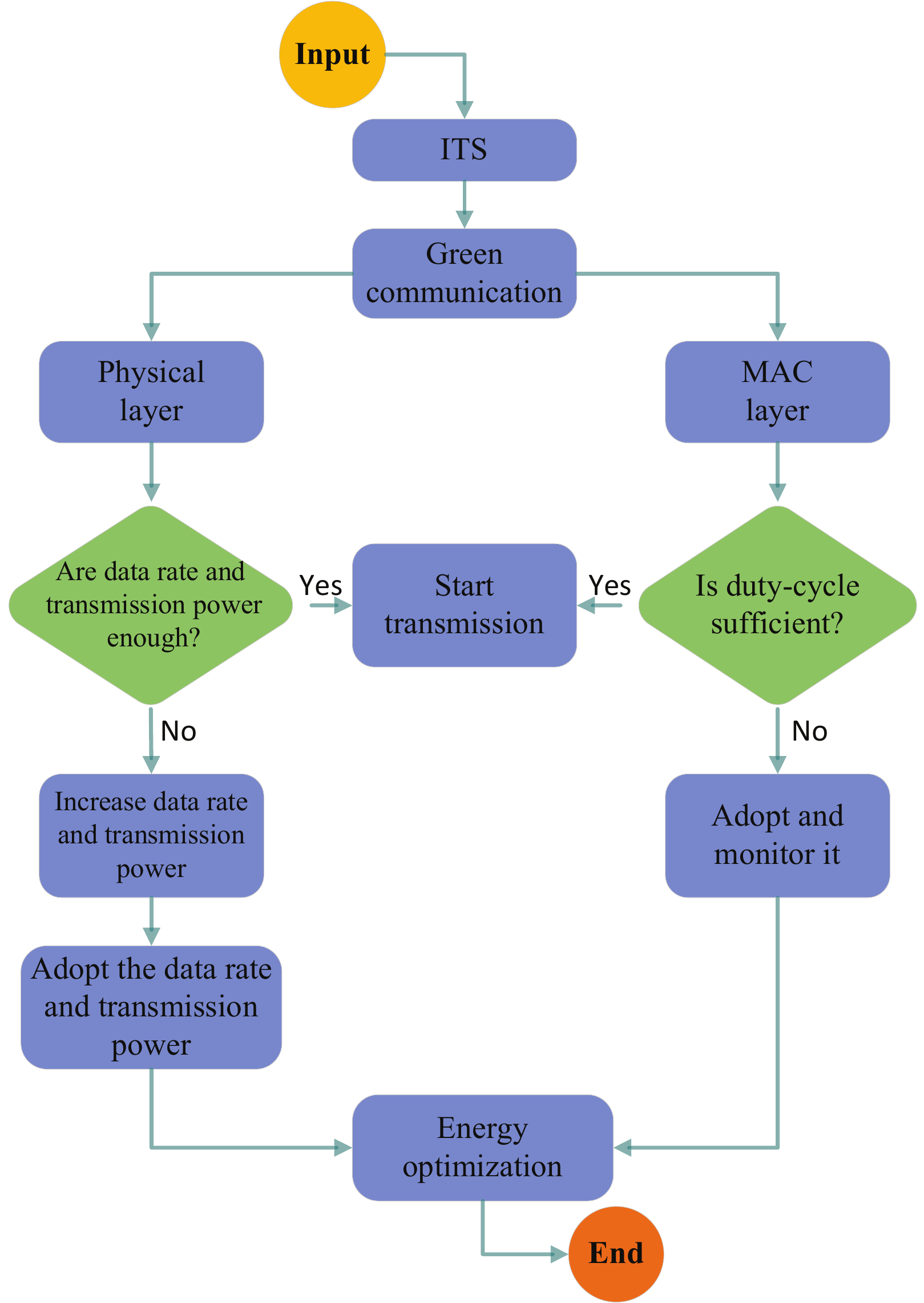}
  \caption{Flowchart of 5G-based green communication algorithm for ITS}\label{fig:communication}
\end{figure}

The above two subsections focused on the V2I and the V2V communication scenarios, respectively. In the IoV, in addition to both the V2I and V2V communications, more sophisticated communications scenarios such as vehicle-to-UAV (V2U), vehicle-to-grid (V2G)\footnote{In a V2G system, plug-in electric vehicles connect to the power grid to let electricity from to/from the vehicles.}, vehicle-to-cloud (V2C), vehicle-to-pedestrian (V2P), and vehicle-to-device (V2D) communications. In this subsection, we focus on the energy-efficient hybrid vehicle-to-everything (V2X) communications where more than two kinds of vehicular communication modes are involved. As shown in Fig. \ref{fig:communication} \cite{Sodhro2020TITS}, an energy-efficiency optimization approach in a vehicular communication framework can consider the parameters related to signaling controlling, data modulation, transmission power adjustment, routing scheduling, node's duty-cycle monitoring, application selection, etc. In the following, we review various green V2X communication techniques in the literature.

\setlength{\textfloatsep}{0.25cm}

\textbf{\emph{1) Optimizing resource allocation:}} Energy-efficient resource allocation has become a prominent direction in achieving green V2X communications. To investigate the power-delay tradeoff in V2X-enabled cellular network, C. Zheng et al. \cite{Zheng2019TVT} propose an energy-efficient relay-assisted transmission scheme based on V2X communications in uplink cellular networks. They derive the theoretical models for both direct transmission mode and V2X-enabled transmission mode considering circuit power and transmit power consumption. A sequential quasiconvex optimization algorithm is first adopted for optimal power allocation without the delay constraint based on which optimal resource allocation is obtained to satisfy the delay constraint and minimize the energy consumption. In \cite{Xiao2020TITS}, the authors consider a cellular network simultaneously supporting multiple energy-harvesting-based D2D links which reuse the downlink resources of cellular users. They formulate a joint optimization problem for resource reuse and power allocation for D2D links in order to maximize the energy-efficiency for cellular D2D-based V2X communication network, with the consideration of energy harvesting constraints and quality-of-service (QoS) constraints.  

With the evolution from 5G to 6G network, more cells will be deployed to provide wide coverage with high data rate, and the vehicle-mounted radio units (RUs) will be promising for future mobile networks. The authors in \cite{Nakayama2020GLOBECOM} propose an adaptive cell zooming scheme for  vehicle-mounted cells, which can adjust the size of each cell according to the current distribution of vehicle-mounted RUs and save energy by reducing the transmission power.

\textbf{\emph{2) Edge caching strategies:}} Novel vehicular networking frameworks have been proposed with the integration of edge caching techniques, which enable vehicles to retrieve contents from nearby edge caching nodes instead of directly requesting contents from the cloud servers. Both energy-efficiency and security are improved in 
edge caching-enabled vehicular networks. T. Limbasiya et al. \cite{Limbasiya2020ACM} propose a vehicular cloud-based secure and energy-efficient communication searching system, which supports secure and effective data storage/extraction in/from the vehicular cloud, as well as the reliable data transmission between different entities. Raja et al. \cite{Raja2021TII} propose to provide green information dissemination services by designing a software-defined vehicular networking (SDVN) framework together with an energy-efficient end-to-end security scheme. Compared with existing frameworks, the proposed SDVN framework can improve the system energy efficiency by reducing the communication and storage overhead.

To make full use of edge caching techniques for reducing communication energy consumption, the content placement optimization has been widely researched \cite{Wu2020JSAC, Wu2020JIOT, Gu2021JIoT, Qiao2020JIoT}. The authors in \cite{Gu2021JIoT} present a framework called cache-enabled satellite-UAV-vehicle integrated network (CSUVIN)  for implementing the popular content distribution among multiple vehicle users (VUs), where a geosynchronous earth orbit (GEO) satellite is regarded as a cloud server and the UAVs are deployed as edge caching servers. Then, the authors propose a coded caching strategy for optimizing the content placement and coded transmission and reducing the backhaul traffic and transmission energy consumption between GEO and UAVs. G. Qiao et al. \cite{Qiao2020JIoT} model the vehicular edge caching optimization problem as a double time-scale Markov decision process, where the content placement decision is made at a large time-scale, and a joint vehicle scheduling and bandwidth allocation scheme is designed on a small time scale. Then, the authors adopt a deep deterministic policy gradient (DDPG) approach to solve a mixed integer linear programming problem to minimize the content access cost.

After the contents have been placed in distributed edge/cloud nodes, the content delivery decision is made to balance the service quality and energy efficiency. In \cite{Xu2018TCSVT}, the authors propose a green information-centric multimedia streaming (GrIMS) framework, which enables on-demand cloud-based processing, adaptive multi-path transmission, and cooperative in-network caching. 
Considering that the 4G/LTE radio can provide long-distance communication at the expense of high energy consumption while the WAVE/802.11p radio supports short-distance communication with a smaller energy consumption, they formulate a joint cost optimal problem and design different heuristic mechanisms for GrIMS network to maximize the QoE (quality of experience) of multimedia transmission while minimizing the network energy consumption.

\textbf{\emph{3) Clustering-based routing:}} In multi-hop vehicular routing, the total transmission distances of all the relaying vehicles can be reduced through optimally selecting certain vehicles as the cluster heads (CHs). With suitable cluster management solutions, the total transmission power consumption can be further reduced by enabling each vehicle to adjust its transmission power as needed \cite{Dong2016INFCOMW}. For the CH selection problem, the authors in \cite{Dua2015DSDIS} propose to select the vehicles with the highest computation, storage capabilities as CHs, and employ the vehicles with high processing capabilities to disseminate data from the information center (i.e., CHs). Furthermore, they use a game theoretic approach to make dissemination decisions based on the current energy situation of the network, which can deal with the energy-consuming broadcast storm problem. The cluster formation in \cite{Kumar2020TITS} includes the selection of the optimized number of control RSUs and the selection of the location of common RSUs, where the control RSUs receive data from the traffic cloud and disseminate to the respective vehicles through common RSUs. Considering the heterogenous traffic data and transmission distance from the smart sensor devices, and probabilistic delay in dynamic vehicular environment, the authors propose the Two-Way Particle Swarm Optimization (TWPSO) algorithm to form optimal clusters for  data routing in Social Internet of Vehicles (SIoV).

\textbf{\emph{4) Optimizing routing paths:}} Due to high dynamicity, uneven distribution and complex topology, it is challenging to design a stable and reliable routing protocol in a vehicular network. On one hand, the unreliable routing increases the network energy consumption due to retransmission operations in failed forwarding. On the other hand, the frequent establishment of routing paths requires to transmit a large number of routing requests (RREQ) and routing reply (RREP) packets, which further consumes a considerable amount of energy. Even though genetic algorithm has been used to search for a set of parameter configurations (i.e., the interval of sending hello messages and topology control message) for the classical energy-efficient OLSR (Optimized Link State Routing) protocol \cite{Toutouh2013CC}, the routing overhead and energy consumption are still high.  To tackle the above problem, X. Wang et al. \cite{Wang2021TGCN} consider the most representative features of vehicles and roadways and use a nonhomogeneous Poisson process to characterize the network connectivity. Then, a fuzzy logic-based routing algorithm is proposed under multiple routing metrics, and show higher routing stability and energy efficiency compared to the classical AODV (Ad hoc On-Demand Distance Vector Routing) algorithm.
For optimizing the route selection decisions, Nash multi-agent Q-learning has been used to train geographic forwarding and traffic data packet replication towards the optimal direction. The objective is to improve the communication energy-efficiency of a vehicular network under the constraints of channel congestion, buffer size, and delay requirement \cite{Zeng2013WN}. The Ant Colony Optimization (ACO) and Particle Swarm Optimization (PSO) algorithm are also applied in optimizing the routing path between the source and destination nodes, by using the nodes' position and speed information \cite{Murugan2020JCR}.

In addition, the energy-efficiency of vehicular routing can be improved through selecting the optimal relaying nodes. Except for the traditional relaying vehicles, various network nodes can be employed as the relays. For example, M. Patra et al. \cite{Patra2017TVT} propose to use mobile Femto Access Points (FAPs) as relays between Macro Base Stations (MBSs) and vehicles, and design the sub-channel power control optimization algorithm to handle the increased co-channel interference between mobile vehicles. The proposed scheme can achieve better delay and energy efficiency performance than the traditional IEEE 802.11p vehicular networks. H. Ghazzai et al. \cite{Ghazzai2019ICVES} propose to use unmanned aerial vehicles (UAVs) as flying relays for delivering data from vehicles to a mobility service center (MSC). In order to minimize the energy consumption of UAVs during the data forwarding and flying process, the authors formulate a routing optimization problem with the consideration of UAVs' residual energy, and design a meta-heuristic particle swarm optimization (PSO) algorithm to decide the selected route and UAVs' positions. Due to the fast movement and frequent changes of driving direction of vehicles, energy-efficient data dissemination becomes challenging in a realistic vehicular network environment. Hence, it is necessary to design novel proactive handover schemes between mobile vehicles and content providers, and enable each vehicle to self-determine the preferred content providers in advance according to real-time positions and routing information. Specifically, the positions of content providers (i.e., surrounded RSUs) can be determined by taking advantage of Doppler effects of periodically received beacon signals \cite{Sun2020TSUSC}. A well-designed handover scheme can reduce the handover overhead, and improve the probability of successful handover and energy efficiency in a vehicular network environment.

\subsection{Summary}
Various green V2X communication techniques have been proposed for different application scenarios. To save the energy consumption of IoV communication infrastructures (i.e., RSUs and BSs),  switching to the sleep mode is an attractive approach. For uplink transmissions from vehicles to IoV infrastructures such as RSUs and BSs, the communication energy consumption can be reduced by selecting energy-efficient relay nodes, deploying edge caching servers, designing highly effective proactive handover schemes, etc. For the multiple routing problems between vehicles, the selected routing path is expected to have higher reliability so as to reduce retransmissions, restrict forwarding hops or choose the optimal forwarding directions so as to reduce transmission bandwidth consumption. In addition, optimizing the resource allocation (i.e., transmission rates and transmission time slots) is appealing to save energy consumption under various V2X communication scenarios.

\section{Green IoV Computation}\label{sec:computation}

\begin{figure}[t]
\centering
  \includegraphics[width=0.49\textwidth]{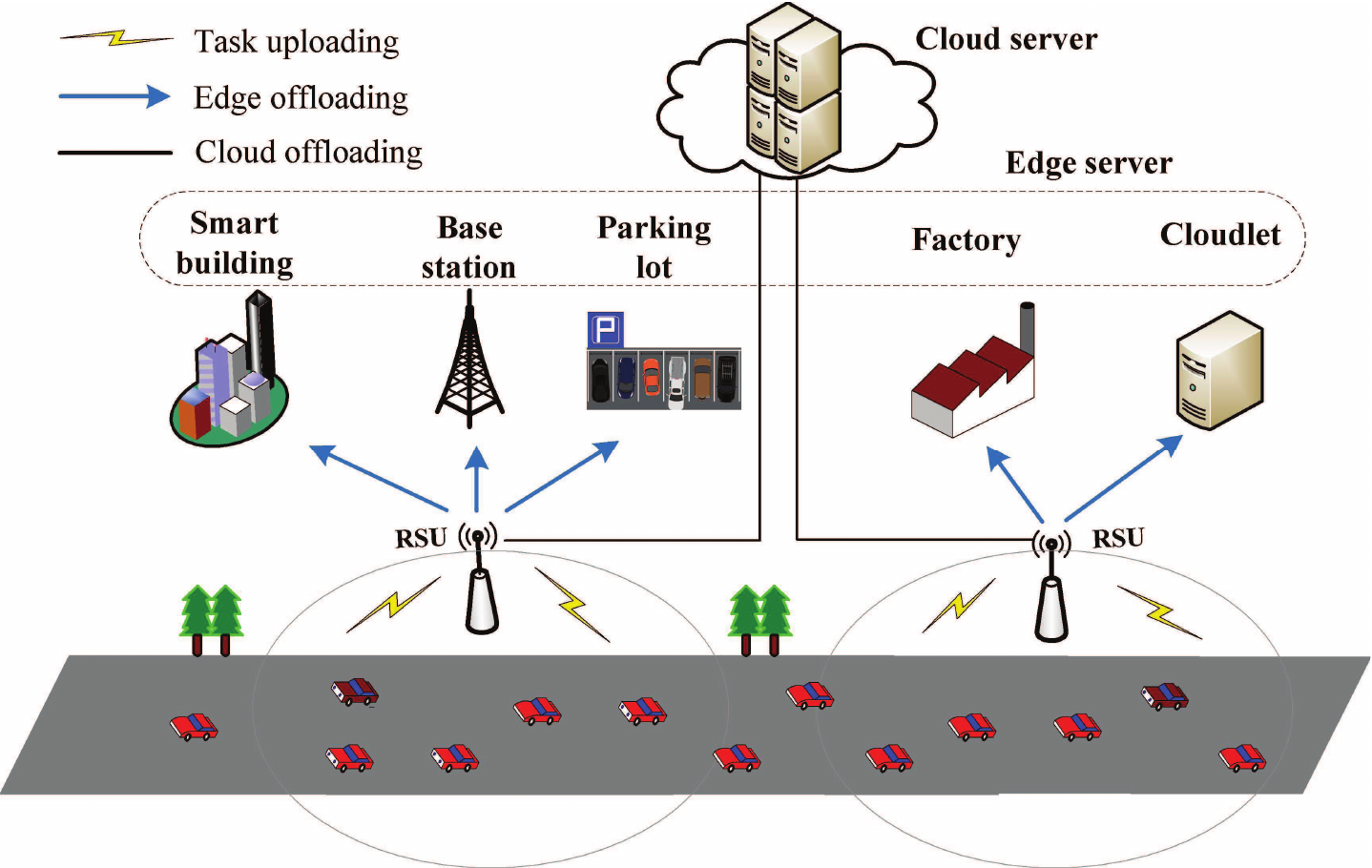}
  \caption{A hierarchical framework for vehicular edge computation offloading}\label{fig:mec}
\end{figure}

Cloud computing and mobile edge computing (MEC) technologies enable mobile vehicles and roadside sensors (e.g., cameras and radars) that have inadequate computation resources to offload tasks to the powerful computing servers for quick responses. Fig. \ref{fig:mec} shows a general multi-layer vehicular edge computation offloading architecture, where a vehicular user can upload tasks to an RSU which will further offload the tasks to other edge computing servers or the cloud server. However, while edge computing provides potential computation resources for IoV, there are challenges in building an integrated and scalable computing architecture with the heterogenous edge devices, balancing workload between the cloud and edge, managing handover, and incorporating idle resources to achieve energy-efficient computing services.

\setlength{\textfloatsep}{0.25cm}

In the following, we first introduce an energy consumption model for vehicular edge computing, and then a survey on the related literatures on green vehicular edge computing, which is summarized in Table \ref{table:computation}.

\begin{table*}[t]
  \centering\caption{Survey of green IoV computation}
  \label{table:computation}
\begin{tabular}{|p{3.1cm}|p{3.6cm}|p{3.4cm}|p{5.6cm}|}
\hline
\textbf{Publication} & \textbf{Offloading scenario} & \textbf{Offloading scheme} & \textbf{Optimization problem or algorithm} \\
\hline
\tabincell{l}{H. Cho, 2020 \cite{Cho2020ICC} \\ M. Liwang, 2021 \cite{Liwang2021TSC} \\ B. Shang, 2021 \cite{Shang2021TVT} \\ Y. Zhu, 2020 \cite{Zhu2020GLOBECOM} \\ M. Shojafar, 2019 \cite{Shojafar2019TCC} \\ Z. Ning, 2019 \cite{Ning2019MNET} \\ Z. Zhou, 2019 \cite{Zhou2019TVT} \\ V. Maio, 2019 \cite{Maio2019ACM} \\ H. Ke, 2020 \cite{Ke2020TVT} \\ Z. Xiao, 2020 \cite{Xiao2020JIOT}} & Vehicle as resource requester & Vehicle-to-MEC offloading & \tabincell{l}{Convex optimization \\ Futures-based resource trading \\ Deep learning \\ Mixed integer convex optimization \\ Gradient-based iterative optimization \\ Heuristic algorithm \\ ADMM optimization \\ Satisfiability modulo theory method \\ Deep reinforcement learning \\ Game theorgy} \\
\hline
\tabincell{l}{A. Alioua, 2018 \cite{Alioua2018WPC} \\ M. Zhu, 2019 \cite{Zhu2019CoRR} \\ L. Zhao, 2021 \cite{Zhao2021TITS} \\ Y. Liu, 2021 \cite{Liu2021TITS}} & Vehicle as resource requester & Vehicle-to-UAV offloading & \tabincell{l}{Sequential game approach \\ Deep reinforcement learning \\ Sequential game approach\\ Successive convex programming}\\
\hline
\tabincell{l}{R. Yadav, 2020 \cite{Yadav2020TVT} \\ Z. Zhou, 2019 \cite{Zhenyu2019TVT} \\ X. Wang, 2020 \cite{Wang2020TMC}} & Vehicle as resource provider & User-to-vehicle offloading & \tabincell{l}{Heuristic approach \\ Pricing-based stable matching \\ Imitation learning based online scheduling} \\
\hline
\tabincell{l}{C. Li, 2019 \cite{Li2019JIOT} \\ T. Bahreini, 2021 \cite{Bahreini2021TMC}} & Vehicle as resource provider & Cooperative vehicular computing & \tabincell{l}{Contract theoretic approach \\ Iterative optimization and greedy algorithms} \\
\hline
\tabincell{l}{R. Meneguette, 2020 \cite{Mene2020TSUSC} \\ L. Pu, 2019 \cite{Pu2019JIOT}  \\ Q. Qi, 2019 \cite{Qi2019TVT}} & Vehicle as resource provider & Integrated vehicular computation offloading & \tabincell{l}{Dynamic virtual machine migration strategy \\ Lyapunov optimization \\ Deep reinforcement learning}\\
\hline
\end{tabular}
\end{table*}
\setlength{\textfloatsep}{0.23cm}

\subsection{Energy Consumption Model}
If a vehicle computes a task locally, the total energy consumption of the vehicle involves only the local computation energy. However, when a vehicle decides to offload the computation task to an edge server, the total amount of consumed energy includes the transmission energy from the vehicle to the edge server and the computation energy at the edge server. Furthermore, if the computation task is dividable, then, the vehicle and the edge server can share the computation workload and both of them consume certain computation energy. For simplicity, we only consider the local energy consumption model (i.e., no offloading) and the offloading energy consumption of indivisible tasks. Assuming that the vehicle transmits a workload with input data size $D_v$ (in bit) to the edge server (i.e., RSU or BS), the transmission energy consumption can be calculated using Eq. (\ref{EqEnergyTrans}). If we denote by $k_v$ (in CPU cycle/bit) the computation-to-volume ratio (CVR), then the local execution time at the vehicle is calculated as:
\begin{equation} \label{EqTimeLoc}
t_v^{loc} = \frac{{{D_v} \cdot {k_v}}}{{{f_v}}},
\end{equation}
where $f_v$ is the local CPU cycle frequency. The local computation energy consumption is calculated as:
\begin{equation} \label{EqEnergyLoc}
E_v^{loc} = p_v^{loc} \cdot t_v^{loc} = \xi  \cdot {\left( {{f_v}} \right)^\beta } \cdot t_v^{loc},
\end{equation}
where $p_v^{loc}$ is the local power consumption which can be modeled by a super-linear function of CPU frequency \cite{Dinh2017TCOMM}. Here, both $\xi$ and $\beta$ are constants.

If the vehicle decides to offload the entire workload to the edge server, the execution time is calculated as \cite{Zhan2020JIOT}:
\begin{equation} \label{EqTimeEdge}
t_e^{exe} = \frac{{{D_v} \cdot {k_v}}}{{f_e^v}},
\end{equation}
where $f_e^v$ is the CPU frequency reserved for the vehicle by the edge server. The power consumption is expressed as \cite{Luo2020JIOT}:
\begin{equation} \label{EqPowerEdge}
p_e^{exe} = \kappa  \cdot {\left( {f_e^v} \right)^\delta },
\end{equation}
where $\kappa$ and $\delta$ are related system parameters of the edge server. Then, the energy consumption of computation offloading is calculated as:
\begin{equation} \label{EqEnergyEdge}
{E_e} = {E_v} + E_e^{exe} = {E_v} + p_e^{exe} \cdot t_e^{exe}.
\end{equation}

In vehicular edge computation offloading problems, the offloading decisions (i.e., local computing or offloading, and where to offload) and workload sharing among the edge servers affect the system energy consumption. In addition, due to the limited resource capacities of vehicles and edge servers, the communication resources allocated for offloading the tasks and the computation resource allocated for processing the tasks are critical factors that affect system energy efficiency.

\subsection{MEC for Vehicular Users}
With the rapid development of vehicular networks, future vehicles are expected to support the intelligent navigation, self-driving, and online gaming, etc. The onboard units (OBUs) may not have adequate computation capabilities to process a large volume of workload with ultra-low latency requirements. Vehicular edge computing (VEC) enables vehicles to offload computation-intensive tasks to network edge servers for saving energy consumption. Various edge servers can be utilized, such as the roadside computing servers that are connected with RSUs/BSs, flying UAVs, etc.

\textbf{\emph{1) Roadside edge servers:}} By offloading tasks to roadside edge servers, the energy consumptions at vehicles are reduced, while transmitting and processing of the offloaded tasks to/at the edge or cloud servers consume extra energy. The network edge servers generally have limited communication, computation and caching capacities, and it is not always the best choice to offload all the computation tasks to the edge servers. Meanwhile, with the deployment of more roadside edge servers to provide communication/computation resources, the construction cost and energy consumption will further increase and need to be considered. Energy-efficient computation offloading strategies will be required for optimal resource allocation and offloading decision for saving energy in vehicular edge computing scenarios \cite{Cho2020ICC, Liwang2021TSC, Zhu2020GLOBECOM, Shojafar2019TCC, Ning2019MNET, Zhou2019TVT, Maio2019ACM, Xiao2020JIOT}.

First of all, optimization of the offloading decision  is a fundamental problem in VEC. To minimize the local energy consumption, each vehicular user chooses either local computing or  task offloading to other edge/cloud servers. However, the edge/cloud servers may not be able to process all the workload in a timely manner due to limited resource capacities, and the system energy-efficiency may be affected. To optimize global energy-efficiency, Z. Zhou et al. \cite{Zhou2019TVT} adopt an ADMM-based distributed solution to decouple the optimization variables (i.e., continuous offloading decisions of vehicles) and decompose the original problem into a series of subproblems for parallel processing. V. D. Maio et al. \cite{Maio2019ACM} develop a satisfiability modulo theories (SMT) method to solve the offloading problem for validation tasks in Blockchain-based VANETs, which aims to balance between energy-efficiency and computation reward. Besides the computation offloading decisions, the communication and computing resource allocations are also critical optimization variables in VEC problems. The authors in \cite{Shojafar2019TCC} propose an energy-efficient adaptive resource management strategy for real-time vehicular cloud services. By exploiting the network states of the TCP/IP connections between fog-enabled RSUs and vehicles, they design a resource management strategy to maximize the overall communication-plus-computing energy-efficiency, through an adaptive control of input/output traffic flows, adaptive reconfiguration of intra-fog communication rates and virtual machine deployments on fog nodes.

Once the computation offloading and resource allocation decisions have been made, cooperation among multiple edge servers is important to further balance the workload and improve computation energy efficiency. In a dynamic traffic environment,  transferring workload between edge/cloud servers will consume more communication energy, and the intermittent connections between vehicles and edge/cloud servers may decrease the QoS and lead to extra energy cost in rebuilding network connections. In \cite{Ning2019MNET}, the authors propose a heuristic algorithm to balance the computation workload among multiple RSUs and minimize the energy consumption due to processing the offloaded tasks under latency constraints. In the considered system, multiple RSUs equipped with MEC servers are connected via fault-free and (almost) delay-free wired links. They periodically broadcast the current network status (i.e., traffic loads and computation capabilities) to other RSUs. When a vehicle uploads the computation tasks to the nearest RSU, it evaluates the service time of these computation tasks based on the incoming traffic flows. If the expected service time is longer than the tolerable delay, the computation task will be transferred to other RSUs via multihop transmission. In the cooperative computation offloading process, the total energy consumption includes the energy consumption in task processing, task transmission among RSUs, and returning the computation results. To minimize the energy consumption of all the computation tasks while satisfying the delay constraint, a MEC-enabled Energy-Efficient Scheduling (MEES) scheme is proposed to schedule computation tasks among MEC servers and feed back the results in an energy-efficient manner to reduce the downlink transmission energy consumption of RSUs.

\textbf{\emph{2) UAV as edge servers:}} The UAV-assisted vehicular edge computation offloading is appealing for enabling intelligent transportation system (ITS) applications. For example, by incorporating UAVs to assist emergency vehicles in data processing, the emergency response time is reduced \cite{Alioua2018WPC}. With the UAVs acting as flying edge servers, the trajectory optimization together with computation offloading and resource allocation becomes the most important issue. M. Zhu et al. \cite{Zhu2019CoRR} use a deep deterministic policy gradient (DDPG) method to jointly adjust the UAV's three-dimensional flight directions, transmission power and channel allocation actions, so as to maximize the total throughput of UAV-to-vehicle communication and improve the system energy efficiency. Besides being edge servers, the UAVs can also act as  relays between vehicles and other roadside edge servers. In \cite{Zhao2021TITS}, the authors present an SDN-enabled UAV-assisted vehicular computation offloading framework to minimize the system cost, where a UAV can act as edge server or a relay between the vehicles and the MEC server. They formulate the offloading problem as a multi-player computation offloading sequential game, and design the UAV-assisted vehicular computation cost optimization (UVCO) algorithm to solve it.

One of the shortcomings of UAVs is their limited battery lifetime, which requires a UAV to fly back periodically to the preset charging stations for  charging. However, with the development of wireless power transmission (WPT) techniques, the UAV can be charged in wirelessly. When a UAV acts as the edge server, it consumes computation resources to process vehicular tasks, and meanwhile, it can be charged using WPT technique. In such a scenario, the resource and energy management becomes complicated. A recently published work \cite{Liu2021TITS} studies a UAV-assisted MEC system for a platoon of WPT-enabled vehicles, where a platoon of vehicles equipped with radio frequency (RF) transmitters can provide energy and offload computation tasks to the UAV. Under the communication and computation resource constraints, the authors incorporate the transmission power of both vehicles and the UAV, the duration of allocated time slots for task offloading (over ground-to-air link), task computing and result downloading (over air-to-ground link) in a joint optimization model. Then, a successive convex approximation-based iterative programming method is proposed to solve the joint scheduling model for improving the overall computation capacity and QoS.

\subsection{MEC With Vehicular  Servers}

With the enhanced computation capabilities of onboard units, mobile vehicles have been employed as dynamic edge computing servers, which can provide computation resources for nearby users (i.e., user-to-vehicle offloading), or relieve the users' computation workload at the BS. Meanwhile, cooperative resource sharing among mobile vehicles (i.e., vehicle-to-vehicle offloading) have been an attractive way to balance the workload and improve network/computation resource utilization.

\textbf{\emph{1) User-to-vehicle offloading:}} When employing the vehicles to provide computation resources for nearby users, large energy cost at vehicles will occur for task processing, and hence, it is not reasonable to assume that the vehicles will share their computation resources unconditionally in practical application scenarios. Suitable incentive mechanisms are required to motive vehicles as edge servers for serving nearby users. In \cite{Zhenyu2019TVT}, the authors propose a contract theory-based incentive mechanism to employ vehicles as edge servers. In the first stage, the BS designs a contract by specifying the relationship between the required resources of users (i.e., performance) and the received rewards of vehicles, where each contract item indicates a distinct performance-reward association. Each vehicle chooses the desired contract item to maximize its payoff. In the second stage, the vehicles that have signed the contract will act as edge servers. The task assignment between vehicles and user equipments is formulated as a two-sided matching problem, and a pricing-based stable matching algorithm is designed to optimize the offloading efficiency.

In different computation offloading scenarios, the energy-delay tradeoff has always been a major research focus \cite{Deng2019TVT, Jang2020VTC}. In \cite{Yadav2020TVT}, the authors propose to exploit the under-utilized computation resources of vehicles to share the workload at cloudlet nodes (i.e., small-size data centers ) during peak times. In order to jointly optimize the energy and latency in task offloading process, a three-phase energy-efficient dynamic computation offloading and resource allocation Scheme (ECOS) is proposed. It first evaluates the resource demands of all the user tasks on the cloudlet nodes and detects the overloaded nodes. Then, the optimal user task which consumes the minimal offloading cost is selected to be offloaded onto a vehicle. Finally, they introduce a heuristic-based joint energy and latency optimization algorithm to discover an efficient vehicle node for processing offloaded tasks. 

Besides classical game theory, contract theory, and traditional optimization algorithms, machine learning (ML)/deep learning (DL)-based AI algorithms including the deep reinforcement learning (DRL) \cite{Ke2020TVT, Dai2020TVT}, federated learning \cite{Xiao2021TITS}, and imitation learning \cite{Nie2021MNET, Wang2020TMC} have been adopted to optimize network performance under dynamic wireless communication environments \cite{Shang2021TVT}. For online task scheduling problems, traditional algorithms mainly use heuristic-based searching ideas, which may have unreliable scheduling performances, especially in highly dynamic vehicular edge computing environments. The authors in \cite{Wang2020TMC} propose an imitation learning-based online task scheduling algorithm for  offloading end users' computation tasks to cooperative vehicles. Imitation learning enables an agent to imitate the expert's demonstrations which are effective solutions for the original problem. Before designing the imitation learning-based task scheduling algorithm, the authors first aggregate the idle resources of service provider vehicles (SPVs) within the RSU's coverage, and model the clustered SPVs as a queueing system during the service time. Then, the online learning policy is obtained by following the expert's demonstration, which is gained by the branch-and-bound algorithm with a few iterations. The proposed scheduling algorithm is shown to result in an acceptable performance gap (with respect to that provided by the expert) in terms of system energy consumption.

\textbf{\emph{2) Cooperative vehicular computing:}} The computation resource coordination among mobile/parked vehicles with well-designed incentive mechanism can improve the energy-efficiency of IoVs  \cite{Li2019JIOT}. The energy savings come from the exploitation of computational slack caused by the discrete computation resource (i.e., CPU/GPU) settings, and the exploitation of nonlinear relationship between these settings and the computation power consumption. For example, when the vehicle system selects a new resource configuration (i.e., activating more CPU/GPU cores or increasing the voltage-frequency level) for completing a larger workload, there may be a computational capacity slack. It may consume more energy when the same-size task is processed by the vehicle operating at a higher CPU frequency instead of the vehicle with lower CPU frequency. Therefore,  energy can be saved by coordinating vehicles under different work modes for completing the offloaded task. T. Bahreini et al. \cite{Bahreini2021TMC} propose an energy-aware resource management framework in vehicular edge computing systems, where energy-efficiency can be improved through sharing and coordinating computing resources among connected EVs. To determine the participating vehicles and resource sharing duration with the unpredictable future locations of vehicles, they design a resource selection algorithm and an energy manager algorithm to select the vehicle state (i.e., requester or provider), the number of workload replications, and the amount of offloaded workload so as to minimize the computation energy consumption of all participating vehicles. When tested under real-world dataset \cite{Uppoor2014TMC}, the cooperative vehicular computing framework shows 7\% to 18\% energy savings compared to that achieved by the local computing, and 13\% energy savings compared to a baseline by offloading all the workloads to RSUs.

\textbf{\emph{3) Hybrid V2X offloading:}} The integrated computation offloading framework enables different types of users (i.e., mobile phone users, vehicle users) to offload tasks to various IoV network nodes (i.e.,  the mobile vehicles, the MEC servers, the cloudlets, the BSs, and the cloud server, etc.) \cite{Xu2019WIRELESS, Mene2020TSUSC, Pu2019JIOT, Wan2021TITS, Qi2019TVT}. These heterogenous vehicular edge computing nodes have different communication and computation capacities, which can provide users with multiple choices. For example, the BS edge server can have larger communication range and powerful computing capability. The fast-moving vehicles may prefer to offload computation-intensive tasks to the BS edge server. On the other hand, the roadside edge server may be more suitable for providing low-speed users such as mobile phone users with high-rate task uploading and relatively quick task processing. In addition, mobile vehicles can form a vehicular cloud and provide on-demand communication and computation resources via virtual machine (VM) migration techniques. When users offload tasks to the integrated computing framework with the mobile vehicular cloud and the cloudlet, the energy expenditure includes the VMs' computation energy in the vehicular cloud and cloudlet and the energy consumed in communications among cloudlets, clouds and vehicles \cite{Mene2020TSUSC}.

Different types of integrated vehicular computing frameworks have been proposed together with various optimization algorithms. The  offloading framework proposed in \cite{Pu2019JIOT} enables vehicular users to leverage the computation resources of cooperative vehicles and the edge cloud. The authors design a Lyapunov 
optimization-based online task scheduling algorithm to minimize the energy consumption of recruited vehicles in heterogeneous crowdsensing applications. S. Wan et al. \cite{Wan2021TITS} propose a 5G-enabled edge computing (EC)-IoV system framework and present the computation offloading scenarios between vehicles, between 5G base stations (gNBs), and between vehicles and gNBs with the assistance of EC servers. The authors design a joint computation task allocation and resource scheduling strategy to balance the workload between gNBs and EC servers and improve the system energy efficiency. Different from the above works, which assume the tasks cannot be divided or the subtasks can be executed independently, the authors in \cite{Qi2019TVT} consider multiple tasks having the data dependence. They present a vehicular edge computation offloading framework with the BSs, cloudlets, and the vehicular nodes acting as edge computing servers. To solve the long-term offloading decision problem, a knowledge driven (KD) service offloading algorithm is proposed, which utilizes a DRL model to learn from previous task offloading and makes decision for the following offloading tasks. Due to the learned offloading knowledge, the KD service offloading algorithm can scruple the data dependence of following tasks, which makes it more suitable for  dynamic online offloading scenarios.

\subsection{Summary}
When powerful computation capacities of edge servers are made available to mobile vehicles, it is important to balance the energy saving at vehicles and the energy consumption in transmitting and processing the vehicular computation tasks. The commonly used edge servers are the fixed edge servers such as the roadside edge servers which can connect to vehicles via APs and RSUs, and the mobile edge servers such as the UAVs. The energy-efficiency can be improved through joint optimization of computation offloading decisions (i.e., local computing or the edge/cloud computing) and resource allocations for communication/computation. Note that, with UAVs as mobile edge servers, the flying trajectories can be optimally determined so as to reduce energy consumption in the flying and the hovering process. In addition, sharing of the computation resources among vehicles can improve the resource utilization and save computation energy of vehicles. Therefore, the main optimization objectives involve VM migration, computation offloading decisions, and resource allocation, etc. The mobility management is another issue, because more energy may be consumed due to transmission interruption and unreturned computation results in dynamic computation offloading environments.

\section{Green Traffic Management}\label{sec:traffic}
In order to make real-time traffic management decisions (i.e., routing decisions and traffic light control), the global traffic information needs to be collected and analyzed. However, it is difficult to retrieve the global information in real time for a large-scale heterogenous and distributed vehicular network. In addition, the traffic management strategies will be frequently updated with the high dynamics of vehicular networks, which is inefficient and energy-consuming. Hence, more researches need to be conducted on the design of low-complexity intelligent algorithms  which enables self-learning and adaptive update in future IoV.

In Table \ref{table:traffic}, we summarize the main approaches for green traffic management. In the following, we first introduce the energy consumption model in different traffic scenarios, and then, compare the existing work on green traffic signal timing, green vehicular routing, driving behavior management, and joint optimization of traffic management.

\begin{table*}[t]
  \centering\caption{Survey of green traffic management approaches}
  \label{table:traffic}
\begin{tabular}{|p{2.6cm}|p{3.2cm}|p{4.5cm}|p{1.4cm}<{\centering}|p{3.5cm}<{\centering}|}
\hline
\textbf{Publication} & \textbf{Optimization algorithm} & \textbf{Related factor} & \textbf{Traffic data collection} & \textbf{Traffic management approach} \\
\hline
J. Zhao, 2016 \cite{Zhao2016TVT} & Heuristic optimization & Vehicle types, speeds, locations and acceleration/deceleration rates, etc. & V2I & Traffic timing optimization \\
\hline
E. Shaghaghi, 2017 \cite{Shaghaghi2017FITEE} & Density/priority-based adaptive traffic timing & Traffic density and queue length of vehicle clusters at intersections & V2X & Traffic timing optimization \\
\hline
L. Chen, 2017 \cite{Chen2017TSMC} & Cooperative intersection control & Car type and length, acceleration habits and turning intention, etc. & V2X & Traffic timing optimization \\
\hline
X. Ge, 2014 \cite{Ge2014TITS} & Heuristic optimization & Value of time (VOT) distribution, route choice and travel demand, etc. &  & Traffic timing optimization \\
\hline
D. Chandramohan, 2020 \cite{Chan2020TESGSE} & IWD optimization & Driving priority & V2I & Congestion avoidance \\
\hline
M. Jabbarpour, 2015 \cite{Jabba2015JNCA} & Ant-based algorithm & Driving directions, vehicle density and speed & V2I & Congestion avoidance \\
\hline
F. Xia, 2018 \cite{Xia2018TII} & Heuristic optimization & Road conditions and traffic information & & Trip generation and traffic assignment \\
\hline
F. Kumbhar, 2021 \cite{Kumbhar2021JIOT} & Machine learning & Vehicle position, speed and direction & V2X & Compatibility time prediction of candidate routes \\
\hline
\tabincell{l}{C. Lin, 2014 \cite{Lin2014ESWA} \\ E. Tirkolaee, 2020 \cite{Tirkolaee2020COIN} \\ Y. Xiao, 2016 \cite{Xiao2016TRE}} & Optimization tool & Traveling distance, road conditions, vehicle types and loads, etc. & & Green traffic-based route planning \\
\hline
\tabincell{l}{G. Mahler, 2014 \cite{Mahler2014TITS} \\ C. Sun, 2020 \cite{Sun2020JIOT} \\ J. Han, 2021 \cite{Han2021LCSYS} \\ S. Bae, 2019 \cite{Bae2019ACC}} & & Traffic timing information & \tabincell{l}{V2X \\C-V2X} & Optimal velocity adjustment \\
\hline
A. Bakibillah, 2019 \cite{Bakibillah2019TVT} &  Bayesian Gaussian learning and stochastic optimization & Historical traffic signal data & Traffic simulator & Optimal velocity adjustment\\
\hline
Q. Lin, 2018 \cite{Lin2018TVT} & Quasi-optimal analysis & & & Optimal velocity adjustment\\
\hline
C. Sun, 2020 \cite{Sun2020JIOT} & Dynamic programming & Historical traffic signal data & V2X & Optimal velocity adjustment\\
\hline
\tabincell{l}{Z. Cao, 2017 \cite{Cao2017TITS} \\ K. Soon, 2019 \cite{Soon2019ASOC} \\ C. Yu, 2018 \cite{Yu2018TRB}} & Optimal and heuristic algorithm & Traffic density, lane choices, etc. & & Coordinated traffic light control and routing planning\\
\hline
\tabincell{l}{C. miao, 2018 \cite{Miao2018TRC} \\ L. Xiao, 2020 \cite{Xiao2020TVT}} & \tabincell{l}{Genetic algorithm \\ Reinforcement learning} & Vehicle speed and position, traffic light information, etc. & V2I & Joint optimization of vehicle trajectory and speed \\
\hline
\end{tabular}
\end{table*}
\setlength{\textfloatsep}{0.23cm}

\subsection{Energy Consumption Model}
By considering different energy efficiency factors under different traffic models, different fuel consumption  models are adopted in the literature. One of the most critical parameters affecting fuel consumption is the vehicle driving speed. By using regression analysis, a speed-type fuel consumption model has been derived  for real-time optimization purpose \cite{Zhao2016TVT}. For example, under steady driving states, the average fuel consumption can be simplified as a function of vehicle driving speed as follows:
\begin{equation} \label{EqF}
F = {k_1} + {k_2} \cdot s + {k_3} \cdot {s^2},
\end{equation}
where $s$ is the vehicle speed, and $k_i$s are the model parameters.

By performing regression analysis, the fuel consumption can be expressed ini terms of vehicle speed, vehicle load and vehicle type. In \cite{Xiao2016TRE}, the adopted fuel consumption rates (FCR) is defined as a function of travel speed $s$ and travel load $f$. For a vehicle type $u$ traveling on a road without slope, the FCR can be calculated by:
\begin{equation} \label{EqFCR}
FC{R_u}\left( {s,f} \right) = {\alpha _u} \cdot {s^{ - 1}} + {\beta _u} \cdot {s^2} + {\gamma _u} + {\varphi _u} \cdot f,
\end{equation}
where $\alpha_u$, $\beta_u$, $\gamma_u$, and $\varphi_u$ are the coefficients related to the vehicle type $u$. The last term of Eq. (\ref{EqFCR}) only works for vehicles with travel loads. The total amount of CO$_2$ emissions during a period is evaluated by multiplying the fuel consumption rates (i.e., $FC{R_u}$) with the fuel-to-CO$_2$ conversion rate of vehicle type $u$ and the traveling distance of the vehicle.

In addition, with the consideration of complex road conditions, the fuel consumption model can be simplified as a function of vehicle speed $s$, road coefficient $\varphi$, and route length $l$ \cite{Miao2018TRC}:
\begin{equation} \label{EqFC}
FC = \sum\limits_{k = 1}^{{M_r}} {f\left( {\varphi \left( k \right),s\left( k \right)} \right) \cdot \frac{{\Delta l\left( k \right)}}{{s\left( k \right)}}},
\end{equation}
where $M_r$ is the total number of sensors located on route $r$, $s\left( k \right)$ is the vehicle speed at sensor location $k$, and $\Delta l\left( k \right)$ is the distance between sensor location $k$ and $k+1$. $\varphi$ is related to the road grade ($\theta$) and rolling resistance coefficient ($\zeta$). The road grade can be calculated by the road altitude and road length information from the global positioning system (GPS):
\begin{equation} \label{EqBeta}
\varphi \left( k \right) = \zeta \left( k \right)\cos \theta \left( k \right) + \sin \theta \left( k \right).
\end{equation}

Without considering the traffic conditions, the fuel consumption of a mobile vehicle is mainly influenced by vehicle-related factors (i.e., types, speed, loads, etc.) and the road-related factors (i.e., length, grade, rolling resistance coefficient, etc.). From the calculation of vehicle fuel consumption, it can be found that by adjusting the traffic signal timing and driving behaviors, vehicle speed can be well controlled, which can save vehicle fuel consumption. Considering the influence of traffic conditions, the traffic signal timing and route selection can reduce fuel consumption through achieving congestion avoidance, reducing traveling delay, and keeping a stable driving, etc.

\subsection{Green Traffic Signal Timing}

Compared to the highway environments, vehicles normally consume more energy under urban traffic conditions due to frequent stopping/starting in traffic jams or at intersections. When the vehicle speed changes quickly, extra energy will be consumed to overcome the static inertia and sliding friction. On the contrary, if the vehicle is driving smoothly,  energy is mainly consumed to overcome rolling friction and wind resistance \cite{USDE2020}. Besides, the traffic signal settings can affect vehicle fuel consumptions and fuel cost through controlling the waiting delays at intersections. Therefore, a well-designed traffic signal control method plays an important role in keeping stable vehicular movements and improving energy consumption at the intersections.

Based on V2X communications, the traffic information at intersections can be aggregated for the traffic manager/controller to make signal timing decisions. The collected traffic data such as the vehicle types, speeds, locations and acceleration/deceleration rates can be utilized to construct the fuel consumption models of different vehicle types \cite{Zhao2016TVT}. Various optimization algorithms (i.e., iterative search, heuristic, learning based approaches) can be designed for determining the traffic signal timing, minimizing the total fuel consumption and traffic delay in passing through the intersection, which are two fundamental optimization objectives in traffic signal timing system. E. Shaghaghi et al. \cite{Shaghaghi2017FITEE} propose a VANET-based traffic signal controlling system (TSCS) to reduce the vehicle waiting time and the pollutant emissions at intersections. The traffic density and prioritized movement information of vehicles are first collected via V2V and V2I communications. Then, the roadside traffic controllers evaluate the intersections' traffic with the vehicles' queue length information, and make traffic signal timing decisions according to the current traffic status near the intersections. Similarly, L.-W. Chen et al. \cite{Chen2017TSMC} propose a cooperative traffic control framework for jointly optimizing the global throughput, the total travel time and average CO$_2$ emissions at multiple intersections. With collected information such as car type and length, acceleration habits and turning intention via V2X communications, the proposed framework enables a cooperative control of queueing sequence and length of traffic lights for adjacent intersections. First, the traffic flow queues are classified into three types according to the state of traffic light, including the stationary queue at the red light, moving queue with stationary vehicles at the beginning of a green light, and the moving queue without stationary vehicles. Then, the adjacent intersections are considered for evaluating the joint passing rate of the whole road network, so as to maximize the global throughput. In addition, the authors further consider an aging process to update the priority of road segment, which guarantees the fairness for each road segment and realizes the green wave driving on arterial roads. Simulation results show that the proposed traffic control framework  improves the global throughput and reduce the average waiting time, total travel time, and average CO$_2$ emission.

\subsection{Green Vehicular Routing}
The green vehicular routing problem (GVRP), which aims at alleviating the road congestions, providing green wave transportation, and further minimizing the CO$_2$ emission and achieving overall energy saving, has attracted a significant amount of attention.

\textbf{\emph{1) Influences of road congestion:}} Traffic congestion causes high fuel consumption and polluted gas emission. The design of vehicle routes can help to avoid obstacles and congestion on the road, release traffic congestion and reduce the overall energy consumptions \cite{Chan2020TESGSE}. M. R. Jabbarpour et al. \cite{Jabba2015JNCA} propose an Ant-based Vehicle Congestion Avoidance System (AVCAS) to reduce fuel consumption and CO$_2$ emissions through the optimal path selection of a vehicle at each time interval. AVCAS includes three main phases. In the initialization phase, the real road map exported from OpenStreetMap is converted to a graph with a set of nodes (i.e., intersections and junctions) and links (i.e., streets and highways). The traffic data such as the IDs and driving directions of vehicles are sent to RSUs for predicting the travel speed of each link. In the second phase, the shortest least congested green paths are constructed between the source and destination based on the collected real-time traffic information. In addition, according to the pheromone value of each link (i.e., related to travel time, length and fuel consumption) and instantaneous states of the vehicle (i.e., density and velocity), the authors compute the probabilities for choosing routes at the intersections. The pheromone value is updated timely, which enables the dynamic selection of the optimal path with reduced traffic congestion and CO$_2$ emissions. Simulation results show that the proposed green VTRS is able to outperform the existing routing approaches in fuel consumption rate by 17\% and alleviate energy consumption effectively.

\textbf{\emph{2) Influences of road condition and vehicle load:}} Except traffic congestion and traveling time, the fuel consumptions and CO$_2$ emissions are also influenced by the road conditions, vehicle loads, driving behaviors, and traffic environment, etc. The urban traffic data (i.e., the road and traffic information) can be utilized to characterize the urban mobility features, based on which, the authors in \cite{Xia2018TII} construct large-scale green urban mobility models and design the trip generation algorithm and traffic assignment techniques for reducing the gasoline consumptions. Similarly, the authors in \cite{Kumbhar2021JIOT} utilize the vehicular mobility information (i.e., position, speed, and direction) available through the vehicular beacon message to estimate the compatibility time (connectivity duration) of two vehicles. Based on an ML model and compatibility analysis, they propose a self-organizing routing protocol, which allows vehicles to estimate or predict the compatibility time of all the candidate routes, and select the best driving route according to the estimation results. The analytical model and five ML techniques including the decision tree (DT), random forest classification (RFC), Gaussian Naive Bayes (GNB), logistic regression (LR), and support vector machine (SVM) are evaluated on the OpenStreetMap (OSM) and SUMO mobility trace generated dataset. Simulation results demonstrate that the proposed scheme can achieve 2 to 3 times higher packet delivery ratio compared to the hop count-based routing algorithms.

It is worth noting that, most of the work on logistics distribution system have focused on the energy-related factors of GVRP such as the traveling distance, road conditions, vehicle types and loads, etc. Researches mainly aim to reduce logistics costs (i.e., the delivery delay, fuel consumption, etc.) by balancing workload between vehicles and selecting the optimal driving routes. From the perspective of supply chain management, the GVRP can be categorized into Green-VRP, Pollution Routing Problem, and VRP in Reverse Logistics \cite{Lin2014ESWA}. In \cite{Tirkolaee2020COIN}, the authors propose a robust green traffic-based routing problem for vehicles to deliver perishable products, which aims to optimize the routing costs (related to the traveling distance), usage costs of vehicles, loading/unloading operation costs, delivery delay, the CO$_2$ emissions and fuel consumption. The problem is relaxed to a deterministic linear model and robust optimization model, which are further validated via CPLEX solver. Similarly, Y. Xiao et al. \cite{Xiao2016TRE} consider to minimize the total CO$_2$ emissions by optimizing the customer-vehicle assignment, route selection and travel time scheduling under time-varying traffic conditions. The constraints include the heterogeneous vehicles (i.e., different vehicle types, CO$_2$ emission models, and time availabilities, etc.), time-varying traffic congestion, customer-vehicle time windows, the vehicle capacity/range, etc. To solve the mixed integer linear programming (MILP) problem, a hybrid algorithm of partial mixed integer programming optimization and iterative neighborhood search (P-MIP-INS) is proposed based on the concepts from variable neighborhood search, which is shown to reduce emission up to additional 8\% on testing data.

\subsection{Driving Behavior Management}

Speed selection according to the traffic signal timing data can assist the vehicle to stably pass through the intersections, so as to reduce the vehicle fuel consumption. With the assistance of V2X communications, the road-side infrastructure (i.e., RSUs or BSs) can broadcast the traffic light cycle information to vehicles approaching the intersection, and then, the vehicles are able to collaboratively adjust their driving speeds and other actions, so as to minimize the total travel delays in the road intersections and reduce vehicles fuel consumption and CO$_2$ emission \cite{Sun2020JIOT, Han2021LCSYS, Kim2020ITSC}. However, when the network infrastructure is not available, the historically averaged timing data and real-time phase data can be used to predict the probability of green traffic lights, based on which, the optimal velocity trajectory can be derived to maximize the chance of passing through green light with improved energy efficiency \cite{Mahler2014TITS}. In \cite{Fayazi2016TITS}, the authors propose to use the crowdsourced probe vehicle data including the GPS coordinates and velocities of the public buses at timestamps, to obtain the deterministic knowledge of signal phase and timing (SPaT) information of traffic lights. In \cite{Bakibillah2019TVT}, a traffic simulator is used to replicate real traffic scenario and measure driving information at the intersections (i.e., signaling, vehicle speed, acceleration and distance to the intersection). The collected information is trained by a Bayesian Gaussian learning model for predicting the traveling delay and passing probability at the intersection. A stochastic optimization algorithm is developed to make the optimal vehicle driving planning.

Different from above works which enable vehicles to adjust driving speed according to traffic signal data, Q. Lin et al. \cite{Lin2018TVT} make the fuel-optimal speed control for vehicles traveling between arbitrary two red-signalized intersections. In addition, since the simplification of powertrain dynamics and fuel consumption model can cause large prediction error in making speed planning at multiple adjacent intersections, the authors adopt more practical fuel consumption models with the consideration of engine, transmission, aerodynamic drag, and rolling resistance, etc. By combining the above factors, the longitudinal behaviours and fuel characteristics in acceleration, deceleration, and constant speed driving can be described more accurately. Based on the fuel consumption model, the authors formulate the speed planning problem as an open-loop optimal control problem, and adopt the Legendre pseudospectral (LPS) technique to solve it. Through analysis, they found that the optimal driving solution between two red-signalized intersections is either a two-stage operation (i.e., accelerating and decelerating) or a three-stage operation (i.e., accelerating, constant speed driving, and decelerating) depending on the distance of the road segment and the vehicle speed limit. A quasi-optimal analytical rule is proposed to approximate the fuel-minimized operating strategy for the red-signalized intersections.

\subsection{Joint Optimization of Traffic Management}

As discussed in the above three subsections, the energy-efficiency of transportation systems can be improved through various approaches, such as the traffic signal management, driving speed control, and route planning, etc. The joint optimization of these traffic control solutions is more promising in realizing a green transportation system. In this subsection, we review the recent works on joint traffic management optimization.

\textbf{\emph{1) Joint optimization of traffic signal timing and trajectories:}} The traffic light control strategy and vehicle routing scheme can be fused and jointly optimized to reduce urban congestion \cite{Cao2017TITS, Soon2019ASOC}. In \cite{Soon2019ASOC}, the authors propose a Pheromone-based Green Transportation System (PGTS) to reduce the GHG emissions and urban congestions. In the first step, they define the `pheromone intensity' to characterize the traffic density of a road network. Based on the pheromone intensity of adjacent upstream road segments, they propose an online epsilon-support vector regression model to forecast the traffic congestion. In the second step, a Coordinated Traffic Light Control (CTLC) strategy is proposed to coordinate the traffic lights on the upstream and downstream of the congested roads, so as to generate a green wave scenario and reduce GHG emissions and fuel consumptions. In the third step, they propose a Cooperative Green Vehicle Routing (CGVR) scheme to probabilistically distribute upstream vehicles to the downstream road, and prevent vehicles entering into the congested upstream road. Through simulations, they prove that the combination of CTLC and CGVR can reduce the frequency of vehicular acceleration when traveling between multiple intersections, which further leads to a decreasing fuel consumption. Similarly, in \cite{Yu2018TRB}, the authors propose an MILP model to jointly optimize vehicle trajectories and traffic signals at an isolated signalized intersection. First, they compute the upper and lower bounds of vehicle arrival times at the stop bars. Second, they formulate an MILP model to minimize the total traveling delay of vehicles passing through the intersection by optimally deciding the phase sequences, duration of each phase, cycle lengths, vehicle lane choices, and vehicle arrival times. Third, they identify the vehicles that pass the stop bars in the same lane as a platoon, and adopt an optimal control model to determine the platoon leading vehicles' trajectories (i.e., acceleration, speed and location) for minimizing total fuel consumption. The proposed model enables all the vehicles to drive through the intersection at desired speeds without stops, and achieve reduced CO$_2$ emissions.

\textbf{\emph{2) Joint optimization of vehicle trajectory and speed:}} In practical traffic scenarios, the vehicle route and speed are coupled in the fuel consumption optimization problems. C. Miao et al. \cite{Miao2018TRC} propose a genetic algorithm and an adaptive real-time scheduling strategy to optimize the vehicular route and speed for improving fuel economy. Here, the fuel consumption model is simplified as a function of vehicle speed, road coefficient and route length. All the related traffic data (i.e., vehicle speed and position, traffic light information) and neighbor vehicle information are collected via V2V, V2I, and V2C communications. On this basis, they model the vehicle macroscopic motion planning (VMMP) problem to minimize the total fuel cost for a given source-destination route. A genetic algorithm-based co-optimization method and an adaptive real-time optimization algorithm are proposed to search the economic route and speed, and update the results at certain frequency, respectively. The proposed algorithms are shown to enhance the vehicular fuel efficiency, for example, up to 15\% fuel economy improvement when compared with the fastest route. Similarly, in \cite{Xiao2020TVT}, the authors formulate a fuel economy optimization problem by jointly designing the traveling route, motion and speed of autonomous driving vehicles (AVs). The AV can receive real-time traffic information (i.e., vehicle velocity and driving direction) from the roadside base stations (RBS) via V2I communications. With collected traffic data, a deep Q-network (DQN)-based algorithm is proposed, which takes the AV as a learning agent to make actions of trajectory selection. The resulting driving safety and fuel consumption are transformed into the agent's reward or penalty. By interacting with the environment, the AV can gradually attain a collision-free and fuel-economic driving policy. In addition, the authors improve the performance of the DQN model by developing a double deep Q-network (DDQN) based algorithm to prevent the over-estimation of the action values. Experimental results show a 24\% fuel economy improvement compared with three other practical driving policies.

\subsection{Summary}
In this section, we have introduced several green traffic management approaches, including traffic signal timing, vehicle routing, driving behaviour control, and joint optimization of these strategies. The optimal decisions are made based on the real-time traffic information such as the vehicle density, driving direction and velocity, which are collected at the intersections via V2I and V2V communications. In the logistics distribution applications, the road conditions, vehicle types, vehicle loads are also important considerations in making green route planning since they affect the  vehicle energy consumption. Except the traditional heuristic algorithm and optimization tool (i.e., CPLEX), the RL-based algorithms have attracted a lot of attention in designing energy-efficient traffic management strategies. The RL-based algorithms can deal with time-varying traffic conditions and make real-time decisions; however, the training time and the learning efficiency of these models are hard to guarantee in dynamic traffic scenarios.

\section{Green Energy Management for EVs}\label{sec:evs}

To make vehicles more environment-friendly, the industry has made great efforts to improve the vehicle design and engineering, such as introducing hybrid power sources (i.e., power from both gasoline engine and the electric motor, or from full electric motor) technology. The emerging EV technologies diversify the `fuels' for the vehicles and are regarded as effective approaches to mitigate the environmental problems. EVs usually include hybrid electric vehicles (HEVs) \cite{kessels2008TVT}, plug-in hybrid electric vehicles (PHEVs), and battery-powered electric vehicles (BEVs).  The power train of HEVs combines a gasoline engine with an electric motor and battery system, which enables HEVs to consume less gasoline and emit less pollution than a traditional internal combustion engine while achieving similar performance \cite{Gallagher2011JEEM}. There are three key electrical technologies for improving the energy-efficiency of HEVs, including the thermoelectric (TE) systems for all hybrids HEVs, the integrated starter-generator (ISG) systems for mild hybrids, and the electronic continuously variable transmission (E-CVT) propulsion systems for full hybrids HEVs \cite{Chau2007JPROC}. Compared with HEVs, the PHEV has an energy storage system as well as an internal combustion engine. The PHEVs are equipped with large battery powers and can be charged as needed, thus further reducing the fuel consumptions \cite{Lin2021TITS}. Different from HEVs and PHEVs, BEVs only use batteries for energy resources and are considered as the future development trend of new energy vehicles. Nowadays, EVs are mostly used to represent BEVs, if not particularly indicated. 


In the following, we first present the general EV energy consumption models and energy consumption optimization approaches. Then, the energy-efficient EV charging techniques, mainly  the EV-to-grid and EV-to-EV charging management optimization are compared.

\subsection{EV Energy Consumption Model}

\begin{figure}[t]
\centering
  \includegraphics[width=0.45\textwidth]{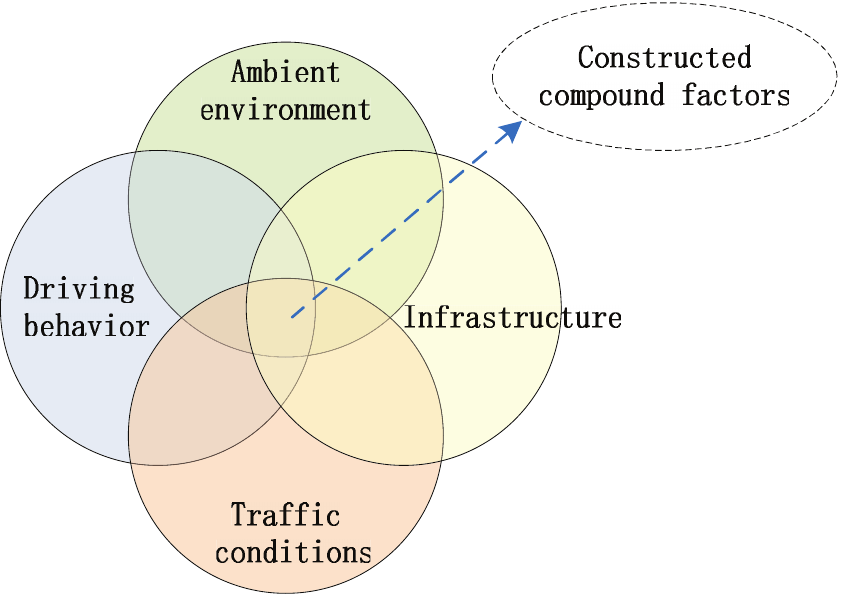}
  \caption{Factors impacting EV energy consumption \cite{Qi2018TRD}}\label{fig:energy}
\end{figure}

The EVs are considered as a promising technology to reduce air pollution and alleviate climate changes. An accurate estimation of EVs' energy consumptions is critical for designing green driving routes and facilitating the development of EVs \cite{Yi2015VPPC, Wang2020APENERGY, Wu2018JCLEPRO, Morlock2020TVT}. In practice, there are various factors that can be influential to the energy consumption of different types of vehicles. Fig. \ref{fig:energy} shows the potential  factors impacting EVs' energy consumption. Except the internal  factors that are associated with the vehicle's powertrain and efficiency parameters, the external factors include the traffic conditions that indirectly influence the vehicle speed and acceleration, the infrastructure-related factors such as grade and surface roughness, ambient environment factors such as the temperature and wind speed, and driving behaviour factors such as driving aggressiveness and driving mode selection, etc. Table \ref{table:model} shows several  EV energy consumption models.

Different from these studies which mainly consider the vehicle speed, acceleration, and road conditions in calculating the fuel consumption \cite{Morlock2020TVT, Wu2018JCLEPRO, Wu2015TRD}, F. Morlock et al. \cite{Morlock2020TVT} propose to forecast the EV energy consumption based on characteristic speed profiles and real-time traffic data which involves information of potential stops at traffic lights and intersections on the trip. A continuous time-dependent speed trajectory is predicted, and based on this, they construct a detailed EV energy consumption model which accounts for specific energy management strategies and environmental factors. Through a field study with Mercedes Benz experimental vehicles, the proposed algorithm is shown to provide an accurate prediction of energy consumption for long look-ahead horizons.
\setlength{\textfloatsep}{0.25cm}

Specifically, the authors in \cite{Qi2018TRD} propose a data-driven decomposition analysis and estimation model for link-level EV energy consumption. Consider a vehicle moving from point A to point B on a road segment with length $L_{link}$ and road grade $\Theta $. Then, the total energy consumption of the vehicle is calculated as:
\begin{equation} \label{EqEC}
{E_{total}} = {E_{tractive}} + {E_{A/C}} + {E_{accessory}},
\end{equation}
where $E_{tractive}$ is the traction energy consumption, $E_{A/C}$ and $E_{accessory}$ are energy consumptions at the air conditioner and other accessories, respectively. The traction energy consumption  is calculated as:
\begin{equation} \label{EqTractive}
{E_{tractive}} = \Delta {E_{kinetic}} + \Delta {E_{potential}} + {E_{rolling}} + {E_{aero}} + {E_{loss}},
\end{equation}
where $\Delta {E_{kinetic}}$ and $\Delta {E_{potential}}$ are the changes of vehicle kinetic energy and potential energy between two points, respectively. $E_{rolling}$ and $E_{aero}$ are the energy consumption in overcoming the road friction and air friction, respectively. $E_{loss}$ is the internal energy loss (i.e., heat loss).

The  energy consumption of a vehicle per unit distance (i.e., energy consumption rate) can be derived by using a regression model, and calculated as:
\begin{equation} \label{EqECR}
ECR \approx \frac{\alpha }{{{L_{link}}}} + \frac{\beta }{2} \cdot m \cdot \left( {PKE + NKE} \right),
\end{equation}
where $\alpha$ and $\beta$ are related coefficients, $PKE$ and $NKE$ are the cumulative positive change in kinetic energy rate and the cumulative negative change in kinetic energy rate, respectively.

To verify the theoretical model, the authors collect the real-world traffic data for energy consumption analysis. Firstly, two impact factors of positive kinetic energy (PKE) and negative kinetic energy (NKE) are analytically proposed. Next, the second-by-second vehicle state (i.e., speed, battery current, air conditioner power, etc.) and road topology (i.e., grade) information are collected from the test vehicle's CAN bus. With data fusion and time-synchronizing techniques, both the GPS data logger and filed data from vehicle data acquisition system are used for analysis. Then, by grouping link-based snippet including roadway type, average speed and grade, etc., the link-level energy consumption rate can be plotted as a linear function of PKE and NKE. In addition, the real-world congestion affects EV energy consumption by affecting the vehicle speed. Finally, three link-level EV energy consumption models are proposed from the perspective of different impact factors. A linear regression fitted model and an artificial neural network (ANN) fitted model are built to characterize these collected traffic data. Experimental results show an improved accuracy of the proposed estimation model over the existing models.

\begin{table*}[t]
  \centering\caption{Classification of EV energy consumption model}
  \label{table:model}
\begin{tabular}{|p{2.8cm}|p{2.8cm}|p{8cm}|}
\hline
\textbf{Publication} & \textbf{Vehicle type} & \textbf{Related factor}\\
\hline
A. Wang, 2020 \cite{Wang2020APENERGY} & Household gasoline vehicles and EVs & Uncertainty in extraction and fuel production, supply share, drive cycles at the same speed, vehicle composition, etc. \\
\hline
Z. Wu, 2018 \cite{Wu2018JCLEPRO} & ICEVs and EVs & Electricity mix, electricity generation technologies, combined heat and power scale, etc. \\
\hline
F. Morlock, 2020 \cite{Morlock2020TVT} & EVs & Traffic factors like vehicle speed, traffic lights on the trip, etc. \\
\hline
X. Qi, 2018 \cite{Qi2018TRD} & EVs & Vehicle speed, battery current, air conditioner power, roadway type, average speed and grade, etc. \\
\hline
X. Wu, 2015 \cite{Wu2015TRD} & EVs & Vehicle velocity, acceleration and roadway grade, etc. \\
\hline
\end{tabular}
\end{table*}
\setlength{\textfloatsep}{0.23cm}

\begin{table*}[t]
  \centering\caption{Energy efficiency optimization for different types of vehicles}
  \label{table:efficiency}
\begin{tabular}{|p{2.4cm}|p{1.4cm}|p{1.9cm}|p{3.6cm}|p{3.6cm}|}
\hline
\textbf{Publication} & \textbf{Vehicle type} & \textbf{Optimization algorithm} & \textbf{Related factor} & \textbf{Optimization variables}\\
\hline
T. Guan, 2016 \cite{Guan2016ITSC} & EVs & DP-based & Travel duration, acceleration and braking, and traffic light conditions, etc. & Velocity and target gear level over a finite horizon \\
\hline
X. Zeng, 2015 \cite{Zeng2015TCST} & HEVs & DP-based & Road grade and speed limit, etc. & Electricity-to-fuel equivalence factor \\
\hline
T. Liu, 2017 \cite{Liu2017TMECH} & HEVs & RL-based & Power demand & Vehicle velocity and power distribution \\
\hline
F. V. Cerna, 2019 \cite{Cerna2019TVT} & HEVs & Linear optimization & Charge-sustaining/charge-depleting modes, SoC of HEVs battery, and deliveries schedules & Driving mode on fuel or electricity, and navigation strategies \\
\hline
G. Ma, 2018, \cite{Ma2018TVT} & HEVs & DP-based & External vehicular dynamics (i.e., position, velocity, etc.) and internal powertrain dynamics (i.e., engine speed, SoC, etc.) & Vehicle level coordination and powertrain level power management \\
\hline
C. Liu, 2020 \cite{Liu2020TNNLS} & PHEVs & Q-Learning and NDP & Future trip information & Power management \\
\hline
J. Liu, 2019 \cite{Liu2019TVT} & PHEVs & HDP-based & Vehicle-speed, engine speed, motor speed, and SoC & Power control \\
\hline
C.-K. Chau, 2017 \cite{Chau2017TITS} & PHEVs & DP-based & Trip information & Drive mode selection and path planning \\
\hline
\tabincell{l}{Q. Zhang, 2020 \cite{Zhang2020TVT} \\ X. Lin, 2021 \cite{Lin2021TITS}} & PHEVs & RL-based & Power demand, vehicle speed, SoC of battery & Route planning and engine power management \\
\hline
\end{tabular}
\end{table*}
\setlength{\textfloatsep}{0.23cm}

\subsection{EV Energy Efficiency Optimization}

Based on different EV energy consumption models, numerous researches have been conducted to optimize the EV's energy efficiency through adjusting the driving speed and driving behaviours with the consideration of current traffic conditions. Table \ref{table:efficiency} summarizes related studies on the energy-efficiency optimization approaches for different types of vehicles.


\textbf{\emph{1) DP-based energy management:}} Dynamic programming (DP) is able to realize the global optimization in a dynamic process, and has been regraded as the most effective approach to solve the energy consumption problems of EVs/HEVs/PHEVs \cite{Chau2017TITS, Ma2018TVT}. T. Guan et al. \cite{Guan2016ITSC} proposed an EV energy-efficiency optimization model for finding a chain of decisions of the optimal velocity and target gear level over a finite optimization horizon, which contains multiple discrete stages. At each stage, the optimization goal is constrained by the maximum velocity and gear, maximum travel duration, acceleration and braking, and traffic light conditions, etc. For solving the problem with complex vehicle model, a stage wise forward-backward Dynamic Programming approach is used to search for the optimal solution to reduce both the total trip time and total energy consumption. Similarly, considering the unknown driving route and road condition ahead, X. Zeng et al. \cite{Zeng2015TCST} model the road grade as a Markov chain and develop stochastic HEV fuel consumption and battery state-of-charge (SoC) models. Then, they formulate the HEV energy management problem as a finite-horizon Markov decision process with the information of the vehicle location, traveling direction and terrain information, etc. A stochastic dynamic programming-based strategy is proposed to improve energy consumption performance by maintaining the battery SoC within its boundaries.

Even though  DP is attractive in designing the EV energy management strategies, it suffers from heavy computation burden and requires future driving information. Hence, other variants of DP such as the adaptive dynamic programming (ADP) and heuristic dynamic programming (HDP) are proposed to offer approximately optimal solutions in online energy optimization problems. The HDP utilizes the nonlinear function fitting to estimate the optimal solution, and adjusts the parameters of nonlinear function through RL to gradually approach the solution of the DP. In \cite{Liu2019TVT}, the authors propose an HDP-based online energy management strategy to optimize the energy consumption of PHEVs. They first adopt the back propagation neural network (BPNN) to construct the dynamic process (i.e., including the data of vehicle-speed, engine speed, motor speed, and SoC) of PHEVs in discrete-time domain. Then, the HDP is used to design an online energy management controller for minimizing the PHEV fuel consumption.

\textbf{\emph{2) Learning-based energy management:}} Learning-based energy management strategies for HEVs/PHEVs, which can operate in different modes to improve fuel economy,  have attracted great attentions. RL is appealing to design model-free adaptive energy management strategy by predicting the related information of HEVs/PHEVs based on historical driving data, such as the vehicles' velocities, driving routes, power demands, driver behaviours, battery lifetimes, etc. \cite{Hu2019MIE, Liu2020TNNLS, Zhang2020TVT, Lin2021TITS}. For example, in \cite{Liu2017TMECH}, the authors adopt the fuzzy encoding and nearest neighbor approaches to predict the vehicle velocity, and exploit a finite-state Markov chain model to learn the transition probabilities of power demand. Then, an RL-based energy management strategy is introduced to determine the optimal control behaviours and power distribution between two energy sources, so as to reduce the fuel consumptions of HEVs. Considering the challenges of energy optimization for PHEVs due to system complexity and physical and operational constraints, C. Liu et al. \cite{Liu2020TNNLS} propose an optimal power management algorithm for PHEVs based on Q-learning and neuro-dynamic programming (NDP), where the evaluated future trip information is incorporated into NDP to make in-vehicle learning possible and effective. Similarly, the authors in \cite{Lin2021TITS} also propose an energy-efficient RL-based power management algorithm for PHEVs, which outputs the engine working power based on the SoC of battery, engine speed, and power demand.

Different from the above work which only focus on the PHEV energy efficiency by optimizing the vehicle speed and powertrain level power management algorithm, the authors in \cite{Chau2017TITS, Zhang2020TVT} consider the relationship between path planning and power management. For instance, given two driving paths with the same length, assume that one path is flat with several traffic lights, and the other has several upper-down slopes with no traffic lights. The time-saving solution may select the second path with no traffic lights, while the energy-saving solution may select the first path if the slopes on the second path lead to more energy consumption. Therefore, the path planning and PHEV power management should be tackled in an integrated framework. Q. Zhang et al. \cite{Zhang2020TVT} propose an RL-based route planning and power management algorithm for PHEVs to reduce energy consumptions. The solution includes inner and outer control loops. The inner loop uses the model-free RL algorithm to output the power control policy according to driving conditions. The outer loop obtains the minimum energy consumption path by minimizing the energy consumption of all the candidate rounds. The effectiveness of the proposed control scheme is demonstrated by simulations in city-wide traffic scenarios.

\begin{table*}[t]
  \centering\caption{Survey of EV charging management}
  \label{table:charging}
\begin{tabular}{|p{3.5cm}|p{2.6cm}|p{7.6cm}|}
\hline
\textbf{Publication} & \textbf{Charging scheme} & \textbf{Optimization approaches} \\
\hline
\tabincell{l}{S. Danish, 2020 \cite{Danish2020TITS} \\ D. Chekired, 2018 \cite{Chekired2018MVT} \\ X. Hu, 2018 \cite{Hu2018MCOM} \\ Y. Cao, 2017 \cite{Cao2017MCOM, Cao2017TVT} \\ L. Cai, 2017 \cite{Lin2017MCOMSTD}} & EV-to-Grid & Optimal selection of charging stations \\
\hline
\tabincell{l}{Y. Zhao, 2020 \cite{Zhao2020TSG} \\ F. Baouche, 2014 \cite{Baouche2014MITS} \\ M. Zeb, 2020 \cite{Zeb2020ACCESS} \\ P. Yi, 2016 \cite{Yi2016TVT} \\ K. Chaudhari, 2018 \cite{Chaudhari2018TII} \\ Z. Sun, 2017 \cite{Sun2017TVT} } & EV-to-Grid & Optimizing deployment of charging stations \\
\hline
\tabincell{l}{H. Kikusato, 2019 \cite{Kikusato2019TSG} \\ Y. Saputra, 2019 \cite{Saputra2019GLOBECOM}} & EV-to-Grid & Coordination among charging stations \\
\hline
\tabincell{l}{M. Shurrab, 2021 \cite{Shurrab2021TITS} \\ R. Zhang, 2019 \cite{Zhang2019TITS} \\ A. Koufakis, 2016 \cite{Koufakis2016SGC}} & EV-to-EV & Optimal matching between EVs \\
\hline
Q. Zhang, 2020 \cite{Zhang2020ITNEC} & EV-to-EV & Optimizing the deployment of flying UAVs \\
\hline
\end{tabular}
\end{table*}
\setlength{\textfloatsep}{0.23cm}

\subsection{EV-to-Grid Charging Management}

Compared to traditional gasoline-powered vehicles, EVs can run out of energy more quickly and need to be charged in a timely manner. Fundamental issues emerge as where to charge the EVs and how to manage the charging process. When the EV is charged in the parking mode, selection of the charging technologies (i.e., normal or fast charging), charging pricing, and utilization of renewable resources  are important in improving charging efficiency. When the EV is charged in the moving mode, the key problems include the designing of EV driving route, the deployment and selection of charging stations. In this paper, we mainly focus on the energy-efficient charging of the EVs, and discuss two types of charging modes, i.e., EV-to-Grid charging and EV-to-EV charging. Table \ref{table:charging} summarizes the related work on EV charging management.

\textbf{\emph{1) Deployment of charging stations:}} The deployment of EV charging stations, which determines where to charge/discharge energy for the EVs, is critical to improve the charging efficiency and energy utilization \cite{Baouche2014MITS}. On one hand, the locations of EV charging stations is of great importance to balance the charging demand and power network stability, as well as improve the charging efficiency and power network capacity \cite{Zhao2020TSG, Sun2017TVT}. On the other head, the economic cost in deploying the EV charging stations should be assessed for realistic application scenarios \cite{Chaudhari2018TII, Zeb2020ACCESS}. In \cite{Yi2016TVT}, the authors introduce a novel architecture, i.e., EV energy internet, to enable the EVs to transmit and distribute energy. An EV that is charged by renewable energy sources can discharge energy at a charging station, from which another EV may be charged. Hence, the renewable energy is distributed in the EV network, which will improve the renewable resource utilization and reduce greenhouse gas emissions. Due to the great importance of charging station placement problem in deploying an efficient EV energy internet, the authors propose two heuristic algorithms (i.e., greedy and diffusion based) to minimize both the number of deployed charging stations and the energy losses. The performances of the algorithms are evaluated with realistic data set of bus systems. Experimental results show that  with the greedy algorithm, it  requires a fewer number of charging stations to cover the whole bus system, and the diffusion-based heuristic leads to less energy transmission losses.

\textbf{\emph{2) Selection of charging stations:}} After the charging stations have been deployed in the EV network, the charging station selection problem should be solved considering the security and privacy of EV users, availability of the reserved time slots, QoS, EV user comfort, and energy efficiency, etc. \cite{Danish2020TITS}. Based on V2X communications, EVs can receive information from nearby charging stations and make energy-efficient charging station selections \cite{Cao2017TVT}. For example, Y. Cao et al. \cite{Cao2017MCOM} propose a communication framework for battery-switch based EV charging. Each charging station is connected to all of the RSUs, and broadcast the condition information (i.e., availability of batteries for switching) to RSUs via V2I communications. On this basis, the EVs can choose the suitable charging stations for battery switch services. The charging station selection logic considers the time for the EV to travel towards the selected charging station and travel from the selected charging station to the EV's trip destination, as well as the time to spend at the selected charging station. Similarly, L. Cai et al. \cite{Lin2017MCOMSTD} propose a V2X communication architecture to facilitate energy-efficient EV charging services in both the charging station scenario and the distributed home charging scenario. With hybrid vehicle communication networks, the EVs can find suitable charging stations and make the best choices according to the potential travel cost, the expected waiting delay and charging cost, etc. Specifically, the real-time communications between the power grid and the EVs enable reliable and energy-efficient charging of a large number of EVs at the same time.

\textbf{\emph{3) Coordination among charging stations:}} For EV networks, prediction of energy demands  is fundamental for coordinating the charging services among charging stations. An SDN-enabled EV charging network architecture can provide a global view of the EV energy demands, the deployment and capacities of charging stations \cite{Wang2018MITS, Hu2018MCOM}. In \cite{Chekired2018MVT}, the authors present a hierarchical software-defined-network (SDN)-based wireless vehicular fog architecture called a hierarchical SDN for vehicular fog (HSVF). The data plane includes data-transmission devices such as RSUs, BSs, and vehicles. The decentralized control plane includes SDN-Fs that are deployed in fog data centers. The SDN control (SDN-C) plane is responsible for constructing the global network connectivity by aggregating information from SDN-Fs and making scheduling decisions (i.e., EV charging and discharging demands) based on the trajectory prediction module, which can mitigate the frequent handover problems with the RSUs and vehicles. Similarly, Y. M. Saputra et al. \cite{Saputra2019GLOBECOM} introduce a centralized communication model for the charging station provider (CSP) to collect information of all the charging stations. A DL-based energy demand learning (EDL) method is developed for the CSP to predict energy demands of the charging stations. Then, to reduce communication overhead and protect data privacy, they further propose to predict the energy demands for EV networks with federated learning, which enables the charging stations to share information without exposing real data sets. The proposed federated energy demand learning (FEDL) approach can improve the prediction accuracy of energy demands up to 24\% and reduce communication overhead by 83\% compared with other baseline ML algorithms.

A cooperative charge/discharge management system can be designed to coordinate energy distribution among multiple charging stations, and even different types of charging stations. Specifically, to reduce CO$_2$ emissions and solve the energy self-sufficiency issue, the concept of net-zero energy houses (ZEHs) is proposed, which aims to realize an annual net energy consumption of zero or less \cite{Bedir2010TDC}. As an important component of ZEHs, the home energy management system (HEMS) maintains the residential convenience by monitoring and controlling the home appliances, energy harvesting, etc. \cite{Mario2015JSAC}. In HEMS, the EVs can be used to minimize the residential operation cost with charge/discharge management schemes. To coordinate the HEMS and grid energy management system (GEMS) for reducing residential operation cost, the authors in \cite{Kikusato2019TSG} propose an EV charge/discharge management framework for effective energy utilization. Based on the voltage constraint information in the power distribution system obtained from GEMS, HEMS determines an EV charging plan to improve the energy-efficiency and minimize operating costs without interfering with EV driving.

\subsection{EV-to-EV Charging Management}

Considering that EVs must be continuously charged due to the low capacity of on-board batteries, and the desire to reduce energy costs, the charging problem of EVs becomes very challenging. Recent works have proposed to investigate more flexible EV-to-EV charging strategies, which can offload the EV charging loads from the electric power systems. With one EV acting as the energy consumer and another EV acting as the energy provider, cooperative energy transfer and sharing are enabled in EV networks. One of the most promising technologies to enable EV-to-EV charging is the wireless power transfer (WPT). Different from traditional wired charging technology, the WPT allows vehicles to be charged while they are moving, which fills the gap of short driving ranges and long charging time of the EVs, and further improves EVs' market penetration ratio and the 
energy-efficiency of EV network \cite{Moon2016TPEL, Sun2018RSER}. Specifically, to improve the customer satisfaction and energy utilization compared the scenario where only EV-to-Grid charging is enabled, the authors in \cite{Koufakis2016SGC} propose an energy scheduling scheme with the EV-to-Grid (V2G) and EV-to-EV energy transfer. They formulate the energy transfer problems as mixed integer programs and solve them offline and optimally in order to improve customer satisfaction and energy utilization. In addition, the authors propose to use a centralized charging station to schedule the offline and online EV charging algorithms, for reducing the charging cost and improving the energy utilization through EV-to-EV energy transfer. They also recommend to introduce a virtual EV to store renewable energy and transfer it to the EVs that is about to arrive at the charging station \cite{Koufakis2020TITS}.

In addition, to make EV-to-EV charging scheduling, a matching based strategy is suitable for selecting EV-to-EV pairs and managing energy transfer among EVs, and further optimizing the users satisfaction and energy-efficiency \cite{Zhang2019TITS, Shurrab2021TITS}. R. Zhang et al. \cite{Zhang2019TITS} propose a flexible EV charging management protocol with EV-to-EV matching mechanism. The EVs as energy consumers and the EVs as energy providers can send their real-time personal information and energy transaction requests to the control center via mobile applications or on-board apps. Then, the control center makes charging/discharging decisions for the involved EVs based on an EV matching algorithm. To achieve an optimal matching, a weighted bipartite graph is constructed to model the charging/discharging cooperation among the EVs, based on which, the maximum weight EV-to-EV matching is obtained. Simulation results verify the efficiency of the proposed EV-to-EV matching based charging strategy in improving network social welfare and reducing EV energy consumption. Similarly, the authors in \cite{Shurrab2021TITS} propose a two-layer matching approach for the EV-to-EV energy sharing problem. A realistic optimization model with user satisfaction, system energy-efficiency, social welfare, and the cost and profit of the EV users is considered. On this basis, they first leverage the Gale-Shapley game \cite{Gale1962AMM} to produce stable matchings among EVs, and then, devise a user-satisfaction model to ensure realistic matchings. The real-life dataset from commercially available EVs is used to test the performances of the algorithms, and the results show the effectiveness and feasibility of the proposed EV-to-EV matching strategy.

\subsection{Summary}
In this section, we have reviewed the development of emerging HEVs, PHEVs, and EVs and different energy consumption models. Various factors may influence the EV energy consumption, including the internal electricity generation technologies and combined heat and power scale, the external traffic conditions such as vehicle speed and acceleration, the road conditions such as the grade and surface roughness, the environmental factors  such as the temperature and wind speed, and driving behaviour factors, etc. Current researches mainly adopt DP-based and RL-based optimization algorithms to reduce energy consumptions of different types of vehicles. For the EVs and HEVs, the vehicle velocity and route become the main optimization variables, while the engine power management has been paid great attention for PHEVs.

In addition, it is important to design novel EV charging management approaches for improving the energy utilization efficiency of an IoV system. For the traditional EV-to-grid charging service, current researches focus on deployment and selection of charging stations, as well as coordination among multiple charging stations for improving charging efficiency. With the development of WPT technology, EVs can be charged on the move and the energy-efficiency can be improved via optimal matching and cooperation among charging/discharging EVs. However, EV-to-EV charging still faces several challenges. For example, there is a lack of related studies on improving the power transmission efficiency and reducing the cost for establishing power transfer system. Besides, it is also challenging to reduce the influence of vehicular mobility on the charging efficiency and achieve seamless switching among different charging facilities.

\section{Energy Harvesting and Sharing}\label{sec:harvesting}

\begin{figure}[t]
\centering
  \includegraphics[width=0.42\textwidth]{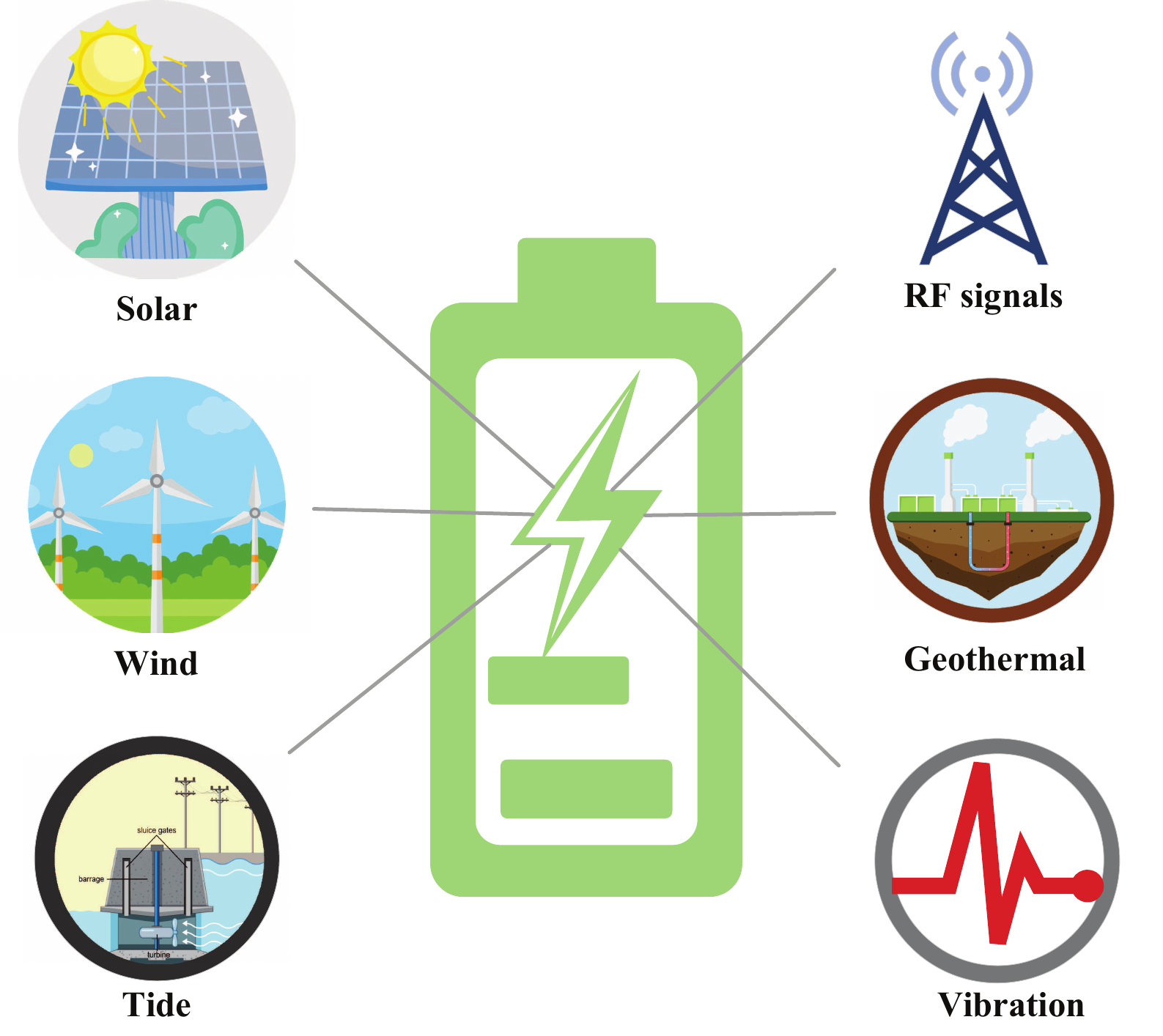}
  \caption{Common energy harvesting sources}\label{fig:renewable}
\end{figure}

Energy supply for the IoV devices and infrastructures is critical to provide  high-quality services. In this context, the energy harvesting technologies will be important which  have received significant  attentions in the research community to deal with the problem of energy constraints in the wireless and IoT devices. Renewable energy harvesting and RF harvesting are two common energy harvesting techniques. Fig. \ref{fig:renewable} shows various energy harvesting sources, including the renewable energy sources such as  solar panels and wind turbines, as well as the RF signal harvesting through which energy in the electromagnetic signals can be converted to electricity for the IoV devices \cite{Mao2021CoRR}. In the following, we first give the energy harvesting model and review the green IoV related researches considering the two energy harvesting techniques.

\subsection{Energy Harvesting Model}

The energy harvesting process can be modeled as successive energy packet arrivals \cite{Mao2016JSAC}. Denote by $E_H^t$ the unit of energy arrival at the mobile mobile (i.e., phones or EVs) at the $t$th time slot, which can be assumed as an independent and identically distributed (i.i.d.) random variable with the maximum value of $E_H^{\max }$. The i.i.d. model is able to characterize the stochastic and intermittent nature of the renewable energy harvesting process \cite{Lak2014JSAC, Mao2016WCNC}. Denote by $e^t$ the part of arrived energy in the $t$th time slot, then,
\begin{equation} \label{EqEHpart}
0 \le {e^t} \le E_H^t,t \in T.
\end{equation}

The energy $e^t$ will be stored in the battery for executing/transmitting the computation task arrived at the following time slots. In the computation offloading applications, the values of $e_t$ can be optimally determined to optimize the battery energy efficiency. Denote the energy consumption of the mobile device at time slot $t$ as $E_m^t$, which consists of the local computation energy consumption and the transmission energy consumption of offloaded tasks. $E_m^t$ satisfies the energy causality constraint:
\begin{equation} \label{EqEHCons}
E_m^t \le {B^t} <  + \infty ,t \in T,
\end{equation}
where $B^t$ is the battery energy level at time slot $t$, and it is constrained by ${B^t} = 0$ and ${B^t} < + \infty ,t \in T, $. Then, the battery energy level at the next time slot $t+1$ can be expressed as:
\begin{equation} \label{EqEHNext}
{B^{t + 1}} = {B^t} - E_m^t + {e^t},t \in T.
\end{equation}

In the computation offloading scenario with energy harvesting enabled mobile devices, the offloading decisions are hard to make due to the varying battery energy level and the coupled relationship of energy consumption/harvesting at successive time slots, and the energy efficiency should be evaluated in the whole scheduling period.

\subsection{Renewable Energy Sources}

Utilizing the renewable energy has been an attractive option for environmental sustainability. While the renewable energy sources can be  used for electronic power generation in a cost-effective way, designing an intelligent energy-harvesting and energy management framework  for IoV is challenging. For example, how to design the energy harvesting and energy management policies for the RSUs according to the prediction of real-time service requests, so as to maximize the energy utilization? Second, for satisfying huge amount of real-time requests, methods for efficient collaboration  among different energy sources deserve to be further studied. Third, since the EVs can store and transfer energy to other EVs and charging stations, it is natural to consider how to design the charging and discharging strategies in realistic traffic environments with varying traffic lights and potential road congestions, etc.

The concept of `energy internet' has been proposed to enable the sharing of renewable energy  (i.e., storing, distributing and controlling) in the information Internet \cite{Hu2018MCOM}. A lot of researches have been conducted in the architecture design of electric power distribution system \cite{Huang2011JPROC},  design and deployment of energy routers (i.e., which can dynamically adjust the energy distribution) \cite{Xu2011SGC, Yi2012ICC}, and the charging/discharging behaviours between energy nodes and vehicles \cite{Wang2019TII}. As battery-powered moving vehicles, the EVs can store and distribute a wide variety of renewable energy when driving in cities. For example, the EVs can store excess electricity such as the solar energy, and return to power grid during on-peak periods through vehicle-to-grid (V2G) connectivity. Then, the other EVs can charge from the power grid and indirectly benefit from the renewable energy \cite{Etezadi2010TPWRD, Yi2016TVT}.

\setlength{\textfloatsep}{0.25cm}

In addition, the renewable energy has also been applied to support other common IoV infrastructures for saving energy, such as the solar-powered RSUs \cite{Atallah2017TVT} and wind-powered RSUs \cite{Muhtar2013ICC}. The conventional RSUs are deployed with connections to the power lines and wired backhauls, which is not feasible in areas with under-developed power systems. Meanwhile, the next generations of ITS applications will require a dense deployment of RSUs to support V2I communication, and their deployment cost will be very high. Therefore, numerous researches have considered to adopt the `green RSU' in future IoV systems. Q. Ali et al. \cite{Ali2016IET} present the design procedure of an RSU armed with the solar energy harvesting circuit and power management module, so as to extend the communication coverage of the IoV infrastructure, and achieve green vehicular communication. With solar-powered RSUs or wind-powered RSUs, to harvest renewable energy as power source, the energy harvesting rate and the RSU density can be further optimized to minimize the network deployment cost \cite{Zhang2017ICC}.

\subsection{Renewable Energy Harvesting}

The renewable energy management mechanism needs to be designed for energy harvesting, storing, distributing between vehicles and IoV infrastructures for improving the energy utilization efficiency. In \cite{Wang2019MWC}, the authors present an energy-harvesting framework with battery-enabled RSUs and EVs for achieving green IoV. RSUs are equipped with wind turbines in one battery cycle, and can sell redundant energy to passing-by EVs via RF energy transfer technology. On the contrary, when the RSU is in mid-level and low-level in battery capacity, and/or the renewable energy generation speed is lower than the energy consumption speed, then the RSU can purchase energy from nearby EVs. On one hand, RSUs intend to maximize the benefits through selling redundant energy to passing-by EVs and minimize the costs of purchasing energy from EVs. On the other hand, EVs hope to minimize the cost of purchasing energy from RSUs and maximize the benefit of selling energy to RSUs. By modeling the requirement processing at the RSU as a queueing system and analyzing the utilities of the RSUs and EVs, the authors further propose a three-stage Stackelberg game-based energy-harvesting strategy, which can balance the utility and energy consumption of the RSUs and EVs.

Furthermore, by adaptive traffic management and energy cooperation based on the energy harvesting level of IoV infrastructures, the energy utilization  can be improved \cite{Abdulla2010GLOBECOM, Lee2020JSYST, Khezrian2015TVT}. With traffic management, the network energy consumption can be reduced through adjusting the served traffic (i.e., by resource allocation) \cite{Zhang2018TVT} and network conditions (i.e., the BS ON/OFF switching, zooming the cell size, and users association, etc.) \cite{Liu2015WCNC} according to the renewable energy level of the IoV infrastructures. The energy cooperation approach can reduce network energy consumption by sharing the renewable energy among IoV infrastructures according to the traffic demands and energy conditions. For instance, the authors in \cite{Khezrian2015TVT} adopt an energy sharing and traffic management approach to achieve green IoV infrastructure. An energy-efficient downlink traffic scheduling strategy is designed for the RSUs when they are empowered by renewable energy source such as solar panel. Then, an online algorithm is developed to allocate RSU and service time slots according to the V2I communication distances and the energy provisioning of RSUs. In \cite{Lee2020JSYST}, the authors design an adaptive traffic management and energy cooperation algorithm to reduce the energy consumption of BSs, which are powered by both on-grid and renewable energy sources. They consider a stochastic system model with uncertain user arrivals, channel conditions and the amount of harvested energy, and adopt a queueing model to manage the traffic demands and energy levels. Then, a Lyapunov optimization-based algorithm is proposed to jointly decide on the energy sharing among BSs, the user association to BSs, and resource allocation in the BSs.

\subsection{Radio Frequency (RF) Harvesting}

Compared to the traditional energy harvesting techniques which utilize the ambient sources such as the solar and wind, the RF energy harvesting is more reliable due to independence on environmental factors. The hybrid access point (HAP) is a kind of RF energy source, which can be configured to support energy transfer and communication demands \cite{Yang2019TGCN, Sangare2015WCNC, Wang2018TGCN}. However, it consumes extra energy when the HAP sends dedicated signals for energy transfer to associated devices. Therefore, energy-efficient resource allocation methods are required for determining the optimal energy supply by the HAP to associated devices \cite{Chen2016TVT, Hadzi2016LCOMM}. The authors in \cite{Abuzainab2017WIOPT} consider a wireless-powered communication network (WPCN) which uses a HAP to serve a set of wireless devices over orthogonal frequency channels. The HAP can not only provide energy to the devices, but can also collect information from devices. The energy signal transmission and information collection are assumed to be over two separate frequency bands. In a WPCN, the HAP can learn the transmission power consumption of the associated wireless transmitter, and further determine the suitable energy signal  for the device. They also propose a robust unsupervised Bayesian learning method to avoid attacks from an adversary who tries to alter the HAP's learning results. The power selection problem for the HAP's energy signal is formulated as a discrete convex optimization problem, and the optimal transmission power is determined from a close-formed solution. Experimental results show that the robust Bayesian learning approach can reduce the packet drop rates without jeopardizing the energy consumption of the HAP. Similarly, J. Kwan et al. \cite{Kwan2020JSEN} also consider a WPCN with several HAPs to collect sensed data from the on-body sensors of a user. Then sensors can temporarily store the harvested energy using storage capacitors. When all the HAPs are out of sensor range, the sensors can only harvest energy from unintended RF energy sources (i.e., nearby mobile phone users). They propose a coordinated ambient/dedicated (CA/D) protocol to optimize the energy harvesting operations of sensors from both the intended RF sources (i.e., HAPs) and the unintended RF sources. The CA/D protocol uses two machine learning techniques (i.e., linear regression-based and ANN-based techniques) to make the optimal energy harvesting decisions with the unpredictable availability of unintended sources and dynamically changing channel conditions between the sensors and unintended RF sources.

There has been considerable interest in utilizing the RF techniques to improve IoV energy efficiency. Since IoV applications require vehicles to periodically upload/download traffic data and scheduling decisions, a large amount of energy is consumed in both transmission and computation processes. Therefore, deploying RF energy harvesting becomes a promising solution for an IoV system to improve energy and spectral efficiency \cite{Mohjazi2015MCOM}. H. Xiao et al. \cite{Xiao2020TITS} consider a radio-frequency-energy-powered cognitive radio network (RF-CRN) for connected vehicles, where the secondary users (SUs) (i.e., vehicles) can harvest energy from RF signals of the primary network (PN) via wireless transmission in the downlink, and then, the SUs can use the harvested energy for uplink data transmissions. Due to co-channel interference among multiple SUs, the authors propose to jointly optimize the transmission time and power control of all the users for maximizing the energy-efficiency of RF-CRN under the given QoS requirement. The non-convex energy optimization problem is transformed into a tight lower-bound convex approximation problem, and the optimal solution is obtained by the Frank-Wolfe (FW) and one-dimensional linear programming approaches.

\subsection{Summary}

Various forms of energy from renewable energy sources can be transformed into electronic power. By equipping the EVs and RSUs with renewable energy harvesting units and designing efficient strategies for energy storage and distribution, the energy-efficiency of IoV systems can be improved. Design of efficient frameworks for energy management, coordination among the energy harvesting sources, energy transfer and cooperative traffic scheduling are critical challenges to be addressed in future IoV research. In addition, the time-varying energy demands of EVs and the dynamic energy supply due to intermittent generation of energy should be jointly considered to improve the energy utilization.

\section{Future Research Trends}\label{sec:future}

\subsection{Green IoV Infrastructures Management}

To provide seamless connectivity for future autonomous driving vehicles, the roadside cameras, radars, RSUs,  MEC servers, and even UAVs will be densely deployed to monitor the real-time traffic conditions, and quickly respond to  emergencies. In addition, the BSs will support V2X communications and provide computation services for mobile vehicles. In the 6G era, the number of BSs is also expected to higher than that in the 5G networks. These IoV infrastructures will consume a huge amount of energy in information collection and processing. Meanwhile, heterogenous hardware, network access, and computing architectures, high dynamicity in the communication environment, and universal connectivity requirement will increase the difficulty in achieving green IoV systems. To realize green IoV service, the deployment and working state adjustment solutions for the IoV infrastructures will need to be well designed. In addition, allocations of the workloads and the resources (for communications and computing)  will be critical to improve user performance and reduce energy consumption. Besides, energy harvesting will be a key technique to reduce the grid electricity demand of the IoV infrastructures.

\subsection{Green Space-Aerial-Terrestrial-Sea Communication}

The space-air-ground integrated networking (SAGIN) will be one of the key technologies in the 6G era. However, the heterogenous hardware platforms, time-varying channel conditions and time-sensitive application requirements complicate the efficient deployment and operation of green space-air-ground communication systems. AI-based channel condition prediction and service requirement prediction, and adaptive online resource allocation techniques will be required in future SAGIN. Current researches mainly focus on the performance optimization of a single layer, such as the joint optimization of flying trajectory and task scheduling between UAVs and vehicles. To improve the overall 
energy-efficiency of IoV systems, cross-layer and end-to-end energy optimization of systems integrating satellites, MEC-enabled
High-Altitude Balloons (HABs), UAVs, and vehicles will be required.

\subsection{Joint Optimization for EV Charging}
Existing researches have proposed a lot of green EV-to-EV and EV-to-grid charging strategies to improve the EV charging efficiency. However,  the existing works lack in joint consideration of the availability and fuel capacities of the charging stations and the EVs, the renewable energy harvesting states, EV driving routes, and even the realistic traffic conditions including the traffic lights and road congestion. How to deal with these complicated factors in EV charging is an open issue. In addition, with the emerging various charging schemes and distributed charging scenarios, AI-based algorithms (i.e., machine learning, state-of-the-art deep learning) for optimizing the charging paths and charging decisions in real time are appealing for future EV charging scenarios. Where to deploy the AI-based algorithms and how to realize quick convergence in real-time scheduling environment and protect the privacy of participants are unsolved issues. Meanwhile, various wireless power transfer technologies can be used for EV charging, such as the resonant beam charging and radio charging. However, the hardware design in improving the charging speed and stability with moving EVs is still very challenging. The cost for establishing the wireless power transfer system will also need to be optimized.

\subsection{Intelligent Energy Harvesting and Sharing}

For the  energy harvesting techniques which are based on the renewable resources such as the solar, winding and tide, it is important to predict the future energy harvesting states and real-time energy demands of nearby IoV devices, so as to make the optimal charging/discharging decisions. In addition, the joint design of intelligent energy-harvesting framework and resource utilization strategies is critical in realizing green IoV scenarios. On one hand, the resource coordination mechanism between IoV devices will facilitate the design of energy harvesting equipments. On the other hand, a well-designed energy-harvesting framework will affect the efficient energy sharing among distributed IoV devices. In order to improve the overall energy-efficiency, mechanisms for efficient collaboration and seamless handover among different energy sources will need to be investigated.

\subsection{Green Heterogenous V2X Communication}

Since V2X communications are supported by heterogenous communication interfaces (i.e., DSRC and C-V2X), it is necessary to develop new protocols which enable seamless handover between various communication interfaces, so as to achieve high-reliability and low-latency communication in a dynamic traffic environment, and improve the energy efficiency. In addition, with a dense deployment of BSs in future 6G networks, frequent switching between the IoV users and the BSs will be unavoidable, and rapidly growing number of IoV users will further increase the energy consumption in communication and computation switching processes. Therefore,  integrated network interfaces and novel communication protocols should be designed to reduce handover cost and improve energy-efficiency for 6G-enabled green IoV systems.

\subsection{Security Issue in Green IoVs}
In the 6G era, more vehicles will be connected to the same IoV infrastructure (i.e., BS or RSU). It will be critical to ensure security in communication and privacy protection among vehicles of different owners. The major security threats come from the unauthenticated users who may cheat the transmitters and receivers by returning misleading information, and achieve malicious access to system resources. In future AI-driven IoV, the adversaries may generate malicious data to mislead the learning direction and cause inefficiency in training performances.

The federated learning  is a decentralized learning framework that utilizes distributed servers to train AI models without uploading data to a centralized server and scheduler~\cite{9460016}. This provides both data security and efficient knowledge sharing. The centralized server only needs the training model parameters from distributed servers, which can save the energy consumption in transmitting data to the central scheduler. In addition, featured by the decentralization, transparency, and immutability, the Blockchain technique can be adopted to ensure trust in an IoV system \cite{Gao2020JIOT}. In blockchain-enabled IoVs, excessive energy may be consumed due to ledger-updates, peer-to-peer smart contracts and transactions. Therefore, how to design an energy-efficient transaction model which can effectively manage the number of transactions and conserve a maximum amount of available energy becomes a critical issue \cite{Sharma2019LCOMM}. Meanwhile, the blockchain technique has been considered for energy trading/sharing scheme for Internet of EVs \cite{Zhou2020TSMC, Sadiq2021ACCESS, Sun2020JIOT}. Different from the traditional centralized power systems, new challenges are emerging in distributed energy trading/sharing environment, for example, how to encourage HEVs/PHEVs/EVs to participate in EV-to-EV energy transactions for improving energy utilization, and design promising consensus mechanism among vehicle to reduce the resource consumption of consensus.

\subsection{AI-based Green IoVs}
AI techniques include the conventional heuristic algorithms, machine learning (ML) algorithms and deep learning (DL) algorithms, etc. Compared to the traditional optimization algorithms which are mainly used in  problems with given assumptions and fixed parameters, AI techniques are able to handle problems with unclear relationships between numerous network parameters, and can be applied in complex scenarios where the solution space is extremely huge and network condition is highly dynamic. In addition, the AI-based algorithms avoid the computation overhead of iterative optimization process, and is more suitable for real-time vehicular network environment. However, the energy consumption of these AI-based methods and algorithms will need to be considered~\cite{9520763}.  By combining the ML/DL models with heuristic-based or game theoretic models (e.g. in a multi-agent scenario), the system energy-efficiency can be improved. In an intelligent traffic management system, AI techniques can be used for adaptive traffic light control with predicted traffic volume at the intersection according to the historical data. In vehicular edge computation offloading scenarios, by predicting the future resource requirements and workload arrivals, the AI techniques can facilitate the switching of the IoV edge servers (i.e., BSs and RSUs) to active/sleep modes, so as to reduce the infrastructure energy consumption. In an energy-harvesting system, AI techniques can track the dynamic energy harvesting states and optimize network configuration by combining the predicted future energy requirements.

\section{Conclusion}\label{sec:conclusion}
We have provided a comprehensive survey on the development of green IoV systems. The  architecture of a green IoV system has been presented and five  scenarios, namely, V2X communication, vehicular edge computing,  intelligent traffic management, electric vehicles and their energy management, and lastly, energy harvesting and management have been focused. For different scenarios of communication, computation, traffic management, electric vehicles,  and energy harvesting management, we have presented the challenges and reviewed the state-of-the-art approaches to achieve energy efficiency. The general energy consumption models for V2X communication, vehicular edge computing, vehicle driving, EV charging and energy harvesting have been also discussed. Finally, we have outlined some open issues in developing 6G-enabled green IoV systems. We anticipate that this article will serve as an important and useful reference in the area of green IoV and spur new researches on this topic.



\ifCLASSOPTIONcaptionsoff
  \newpage
\fi

\bibliographystyle{IEEEtran}
\bibliography{IEEEbib}

\bibliographystyle{IEEEtran}
\vspace{-10 mm}
\begin{IEEEbiography}[{\includegraphics[width=1in,height=1.25in,clip,keepaspectratio]{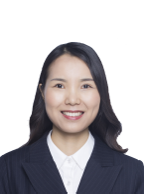}}]{Junhua Wang}
received the BS degree and PhD degree in computer science from Chongqing University, in 2014, and 2019, respectively. From 2017 to 2018, she was a visiting scholar in University of Houston, USA. Since 2019, she has been working with the College of Computer Science and Technology, Nanjing University of Aeronautics and Astronautics. Her research interests include mobile computing, vehicular ad-hoc networks, and wireless networks.
\end{IEEEbiography}

\vspace{-10 mm}
\begin{IEEEbiography}[{\includegraphics[width=1in,height=1.25in,clip,keepaspectratio]{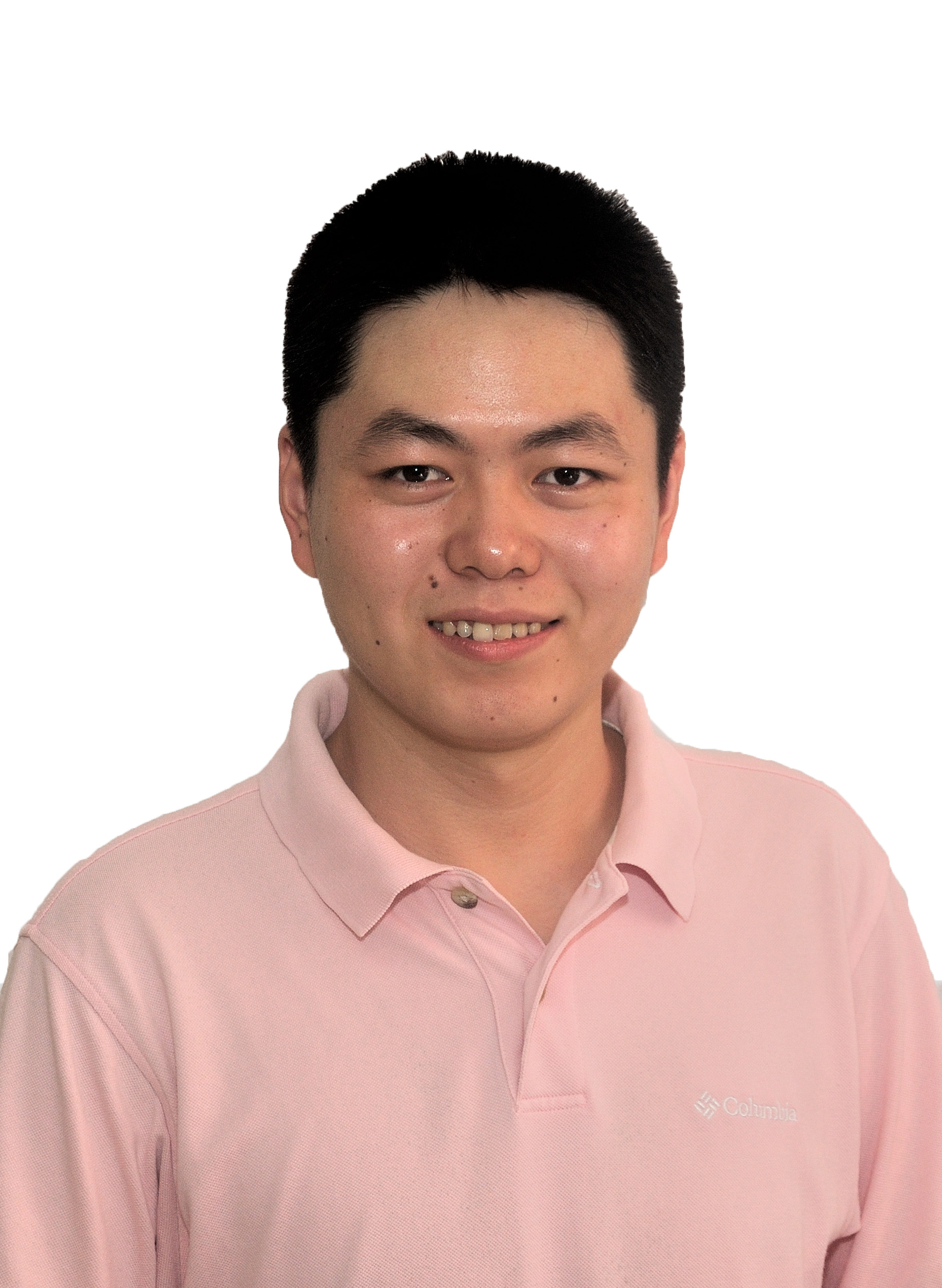}}]{Kun Zhu (M'16)}
is currently a Professor in the College of Computer Science and Technology, Nanjing University of Aeronautics and Astronautics, and the Collaborative Innovation Center of Novel Software Technology and Industrialization, Nanjing, China. He is also a Jiangsu specially appointed professor. He received his Ph.D. degree in 2012 from School of Computer Engineering, Nanyang Technological University, Singapore. His research interests include resource allocation in 5G, wireless virtualization, and self-organizing networks. He has published more than fifty technical papers and has served as TPC for several conferences. He won several research awards including IEEE WCNC 2019 Best paper awards, ACM China rising star chapter award.
\end{IEEEbiography}
\vspace{-10 mm}
\begin{IEEEbiography}[{\includegraphics[width=1in,height=1.25in,clip,keepaspectratio]{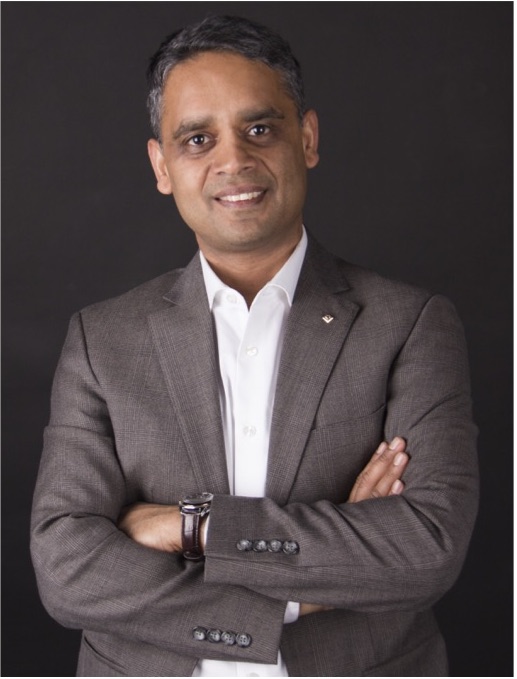}}]{Ekram Hossain (F'15)}
is currently a Professor and the Associate Head (Graduate Studies) in the Department of Electrical and Computer Engineering at University of Manitoba, Canada. He is a member of the College of the Royal Society of Canada (Class of 2016), a Fellow of the Canadian Academy of Engineering, and a Fellow of the Engineering Institute of Canada. His research interests include design, analysis, and optimization of wireless and mobile communication networks and applied machine learning for intelligent wireless connectivity with emphasis on beyond 5G cellular networks. He was elevated to an IEEE Fellow for contributions to spectrum management and resource allocation in cognitive and cellular radio networks. He was listed as a Clarivate Analytics Highly Cited Researcher in Computer Science in 2017, 2018, 2019, and 2020. He was an elected member of the Board of Governors of the IEEE Communications Society (2018-2020). He received the 2017 IEEE ComSoc TCGCC (Technical Committee on Green Communications \& Computing) Distinguished Technical Achievement Recognition Award ``for the Outstanding Technical Leadership and Achievement in Green Wireless Communications and Networking".  He currently serves as the Editor-in-Chief of the IEEE Press (2018-2021). Previously, he served as the Editor-in-Chief for the IEEE COMMUNICATIONS SURVEYS AND TUTORIALS (2012-2016).
\end{IEEEbiography}

\end{document}